\documentclass[twocolumn]{aastex631}

\usepackage[utf8]{inputenc}

\usepackage{graphicx}
\usepackage{txfonts}
\usepackage{lipsum}
\usepackage[T1]{fontenc}

\shorttitle{VLA Radio Survey of Magnetic Activity in 140 Exoplanets}
\shortauthors{Ortiz Ceballos et al.}

\graphicspath{{./}{figures/}}

\begin{document}

\title{A Volume-Limited Radio Search for Magnetic Activity in 140 Exoplanets with the Very Large Array}

\author{Kevin N. Ortiz Ceballos}
\affiliation{Center for Astrophysics ${\rm \mid}$ Harvard {\rm \&} Smithsonian, 60 Garden St, Cambridge, MA 02138, USA}

\author{Yvette Cendes}
\affiliation{Center for Astrophysics ${\rm \mid}$ Harvard {\rm \&} Smithsonian, 60 Garden St, Cambridge, MA 02138, USA}
\affiliation{Department of Physics, University of Oregon, Eugene, OR 97403, USA}

\author{Edo Berger}
\affiliation{Center for Astrophysics ${\rm \mid}$ Harvard {\rm \&} Smithsonian, 60 Garden St, Cambridge, MA 02138, USA}

\author{Peter K. G. Williams}
\affiliation{Center for Astrophysics ${\rm \mid}$ Harvard {\rm \&} Smithsonian, 60 Garden St, Cambridge, MA 02138, USA}

\begin{abstract} 
We present results from a search for radio emission in 77 stellar systems hosting 140 exoplanets, predominantly within 17.5 pc using the Very Large Array (VLA) at $4-8$ GHz. This is the largest and most sensitive search to date for radio emission in exoplanetary systems in the GHz frequency range. We obtained new observations of 58 systems, and analyzed archival observations of an additional 19 systems. Our choice of frequency and volume limit are motivated by radio detections of ultracool dwarfs (UCDs), including T dwarfs with masses at the exoplanet threshold of $\sim\!13\,M_J$. Our surveyed exoplanets span a mass range of $\approx\,10^{-3}-10\,M_J$ and semi-major axes of $\approx\,10^{-2}-10\,$AU. We detect a single target -- GJ\,3323 (M4) hosting two exoplanets with minimum masses of 2 and 2.3$\,M_\oplus$ -- with a circular polarization fraction of $\approx\,40\%$; the radio luminosity agrees with its known X-ray luminosity and the G\"udel-Benz relation for stellar activity suggesting a likely stellar origin, but the high circular polarization fraction may also be indicative of star-planet interaction. For the remaining sources our $3\sigma$ upper limits are generally $L_\nu\lesssim\,10^{12.5}\,\mathrm{erg}\,\mathrm{s}^{-1}\,\mathrm{Hz}^{-1}$, comparable to the lowest radio luminosities in UCDs. Our results are consistent with previous targeted searches of individual systems at GHz frequencies while greatly expanding the sample size. Our sensitivity is comparable to predicted fluxes for some systems considered candidates for detectable star-planet interaction. Observations with future instruments such as the Square Kilometer Array and Next Generation Very Large Array will be necessary to further constrain emission mechanisms from exoplanet systems at GHz frequencies.
\end{abstract}


\section{Introduction} 
\label{sec:intro}

Observational constraints on the magnetic activity of exoplanets are extremely limited. While the magnetic fields of all magnetized solar system planets have been measured directly via astronomical observations or in-situ measurements \citep{stevenson_planetary_2003}, no confirmed direct detection of a magnetic field has been achieved for an exoplanet.
Several techniques exist for indirectly estimating the magnetic field strength of exoplanets. Observations of star-planet interactions have been used to constrain exoplanet magnetic fields, for example by identifying modulations in \ion{Ca}{2} chromospheric emission from the star in phase with the planetary orbit \citep{shkolnik_evidence_2003, shkolnik_hot_2005, gurdemir_planet-induced_2012, cauley_magnetic_2019}, as well as periodic X-ray emission in phase with the orbital period \citep{acharya_x-ray_2023}. Transit observation of atmospheric bow shocks \citep{cauley_magnetic_2019} and evaporating atmospheres \citep{ben-jaffel_signatures_2021, schreyer_using_2023} have also been used to estimate planetary magnetic fields.
However, these methods are indirect and offer uncertain estimates at best.

In the solar system, radio observations serve as direct probes of the magnetic fields of the giant planets \citep{burke_observations_1955, zarka_ground-based_1997}. The solar system planets emit radiation at radio frequencies through the Electron Cyclotron Maser Instability (ECMI) mechanism, which causes emission up to a maximum frequency directly proportional to the maximum magnetic field strength \citep{zarka_auroral_1998}. The nonthermal, incoherent gyrosynchrotron process is also present in Jupiter's radio emission, but it is a much weaker signature due to its inefficiency \citep{zarka_magnetospheric_2015}, making ECMI measurements the strongest diagnostic of planetary magnetic field in the solar system.

Searches for radio emission from exoplanet systems, across MHz to GHz frequencies, have so far yielded non-detections \citep[e.g., ][]{winglee_search_1986, zarka_ground-based_1997, bastian_search_2000, lazio_radiometric_2004, lazio_magnetospheric_2007, lazio_blind_2009, lynch_search_2017, ogorman_search_2018, route_rise_2019, cendes_pilot_2021, route_rome_2023} or tentative detections \citep[e.g., ][]{lecavelier_des_etangs_gmrt_2011,lecavelier_des_etangs_hint_2013}.
In general, the detection of stellar emission at radio frequencies is still challenging. While the very closest stars are sometimes detectable in their thermal emission \citep[e.g. $\alpha$ Centauri;][]{trigilio_detection_2018}, these are exceptions due to their extremely close distances. Rather, stars are often observable in the radio due to non-thermal emission, such as cyclotron masers and gyrosynchrotron radiation \citep{dulk_radio_1985}, a variable type of emission found across a large portion of the radio spectrum \citep{hughes_unlocking_2021}.
Recently, non-targeted searches through source location cross-matching on radio sky surveys have permitted new discoveries of radio-bright main-sequence stars at MHz \citep{callingham_population_2021, gloudemans_plausible_2023} and GHz \citep{driessen_detection_2023} frequencies. However, there is yet no evidence that these signals are definitively tied to exoplanets in these systems. A recent promising detection of flaring $2-4$ GHz radio emission from YZ Ceti, which hosts a short-period planet, may be co-periodic with the planet's orbit, potentially indicating star-planet interaction \citep{pineda_coherent_2023}. 

Searches that have sought to find emission directly from exoplanets (as opposed to from star-planet interactions) have more recently focused on the MHz  regime. Jupiter's ECMI emission, caused by its 14 G magnetic field, reaches a maximum cyclotron frequency of about 40 MHz \citep{zarka_planetary_2012}. An exoplanet with a magnetic field similar to Jupiter, or up to a few times stronger, would still emit at tens or hundreds of MHz. Two results in this regime have so far been presented as tentative detections. A potential signal from the Tau Bootis system \citep{turner_search_2021} was detected with LOFAR, but was seen only once and could not be ruled out as being of stellar origin; follow-up observations showed no sign of emission \citep{turner_follow-up_2024}.
Another signal, from the direction of GJ 1151, has also been reported from LOFAR data \citep{vedantham_coherent_2020}, but follow-up radial-velocity measurements rule out the presence of a Jupiter-mass companion \citep{pope_no_2020}. Later observations revealed a long-period ($P = 390$ d) exoplanet, likely too low mass ($M_p \sin i = 10.62$ M$_\oplus$) to be the source of the signal \citep{blanco-pozo_carmenes_2023}. Further LOFAR detections of circular polarization in a subset of M dwarfs have been likewise attributed to exoplanet interactions \citep{callingham_population_2021}, although all but two of these newly detected sources are not known to host exoplanets. 

On the other hand, GHz frequency radio observations of very low mass stars and brown dwarfs (hereafter, ultracool dwarfs, UCDs) have proved fruitful \citep[e.g.][]{berger_discovery_2001,berger_flaring_2002, hallinan_periodic_2007, route_arecibo_2012,mclean_radio_2012, route_second_2016, kao_auroral_2016}. Over two dozen brown dwarfs with spectral types L and T, have been detected in the radio \citep{berger_flaring_2002,mclean_radio_2012, williams_radio_2018, kao_radio_2022}.
The detection of emission from the T2.5 dwarf SIMP J01365663+0933473 ($M = 12.7 \pm 1.0$ M$_J$) established that even planetary-mass objects can emit at GHz frequencies \citep{kao_strongest_2018}.
Unlike the magnetic field of a star like our Sun, which is generated by shear in the tachocline \citep{parker_hydromagnetic_1955}, the dynamos of UCDs are thought to be convection-generated, which is also the case for planets in our solar system \citep{christensen_energy_2009}. 
This dynamo process was initially predicted to generate only weak magnetic fields, but this has now been refuted by the properties of the radio emission, which require kG-level large-scale fields \citep{berger_flaring_2002,williams_trends_2014,hallinan_magnetospherically_2015}. In fact, recent results have shown spatially resolved emission around the UCD LSR J1835+3259, which potentially indicates the presence of a planet-like radiation belt \citep{kao_resolved_2023, climent_evidence_2023}, suggesting that the strong magnetic fields in UCDs may be ``planet-like'' in nature \citep{williams_radio_2018}.
The detection of GHz frequency radio emission from UCDs thus implies that exoplanets may also be capable of generating strong enough magnetic fields to cause detectable radio emission at these frequencies, where sensitive searches can be carried out.  This serves as the main motivation for this work.

In \citet{cendes_pilot_2021}, we conducted a pilot search for GHz frequency emission from a small sample of five systems with eight exoplanets, which had all been discovered via direct imaging. Directly-imaged exoplanets are an attractive sample due to their comparable mass scale to T dwarfs, and due to their resolvable angular separation from their host stars in the VLA observations.
Furthermore, these planets are generally younger and warmer, and thus expected to have stronger convection and a more active dynamo \citep{reiners_magnetic_2010}. Our pilot study did not detect any of these targets, but established luminosity upper limits of $\lesssim 10^{12.5}$ erg s$^{-1}$ Hz$^{-1}$, comparable to the detected emission from some T dwarfs \citep{pineda_panchromatic_2017}.

The number of nearby directly-imaged exoplanets is currently small, especially in the context of radio detection rates of UCDs of $\sim 5-10\%$ \citep{berger_flaring_2002,mclean_radio_2012,route_second_2016}. To achieve statistically meaningful results that could constrain the presence of radio emission from exoplanet systems requires a much larger sample of nearby systems.  Such a sample will also naturally span a wide range of masses, thereby exploring radio emission from Earth-mass to multi-Jupiter mass systems. Here, we present the results of the first large-scale GHz-frequency survey of nearby exoplanet systems, predominantly within 17.5 pc using the Very Large Array (VLA), combining new data with archival observations. In \S\ref{sec:survey} we present the survey and experimental design. In \S\ref{sec:results} we present the results of the observations, and in \S\ref{sec:discussion} we discuss their implications; we end with concluding remarks in \S\ref{sec:conclusions}.

\begin{table*}[t!]
    \hspace{-1.5cm}
	\centering
	\begin{tabular}{lllllll}
	\toprule
	
	Program ID & Dates Observed & Configurations & Targets Observed & Targets Used \\ 
	
	\hline

	22A-186 & 2022-03-01 to 2022-07-02 & A, BnA$\rightarrow$A & 37 & 35\\
	23A-270 & 2023-03-29  to 2023-05-14  & B & 23 & 23\\
	\hline
	15B-326  & 2015-11-17 to 2016-01-21 & D, DnC & 21 & 5\\
	18B-048 & 2019-01-14 to 2019-02-16 & C, C$\rightarrow$B & 27\footnote{The number of targets observed in C-band for this program.} &  14\\

	\hline\hline
	\end{tabular}
	
	\caption{VLA programs used in this study.}
    \label{tab:programs}
	
\end{table*}

\vfill\null

\section{Sample Selection and Observations} 
\label{sec:survey}

We constructed a target sample using the NASA Exoplanet Archive\footnote{\href{https://exoplanetarchive.ipac.caltech.edu/}{https://exoplanetarchive.ipac.caltech.edu/}}, which included about 5,500 confirmed exoplanets at the time of the sample construction in early 2023.
We imposed the following selection criteria: (i) companion mass of $<13$ M$_J$ to ensure exoplanet targets; (ii) distance of $<17.5$ pc to ensure that we can reach luminosity limits of about $10^{12.5}$ erg s$^{-1}$ Hz$^{-1}$, comparable to the faintest UCDs, in a reasonable amount of observing time; and (iii) declination of $>-25^{\circ}$ for accessibility and ease of scheduling with the Very Large Array (VLA). 
This led to a complete, volume-limited target sample of 83 targets containing 145 exoplanets.
Of these targets, we conducted new observations of 58 targets
\footnote{One additional target, 61 Vir, was also observed in our program but its location was contaminated by bright emission from a nearby source; we therefore consider it an unobserved target.}.
We further supplemented these observations with analysis of archival data for an additional 12 targets.
In total, we present results for 70 of the 83 targets in this first target sample.
In addition, we also include in our survey 7 targets that are beyond the 17.5 pc cutoff: One target (70 Vir) for which we obtained new observations, and 6 targets that were included in the archival datasets we analyzed. A summary of the number of targets observed, and the number of targets used in the results of this study, is provided in Table~\ref{tab:programs}.

\subsection{New VLA Observations} \label{sec:observations}

We obtained observations with the VLA as part of programs 22A-186 (PI: Cendes) and 23A-270 (PI: Ortiz Ceballos); details are shown in Table~\ref{tab:programs}. All observations were performed in the C-band, with continuous spectral coverage at $4-8$ GHz. We selected C-band due to its optimal sensitivity, and since UCD radio emission has been predominantly detected at this frequency range \citep[e.g.][]{berger_magnetic_2005, berger_periodic_2009, williams_quasi-quiescent_2013, kao_auroral_2016, kao_strongest_2018}. We selected observing times proportional to the distance to each target to achieve a luminosity limit of $\approx 10^{12.5}$ erg s$^{-1}$ Hz$^{-1}$ or better across the sample.

\vfill\null

\subsection{Archival Data}

We additionally identified unpublished data in the VLA archive that include exoplanets within our 17.5 pc cutoff (or close to it). These programs are listed in Table~\ref{tab:programs}, along with their observational details. For program 18B-048 (PI: Bastian), we only used observations in C-band to match our own data. In the case of both archival programs, we also excluded targets that we already observed as part of our new programs, given our greater sensitivity.

\subsection{Data Analysis}
\label{sec:analysis}

\begin{figure*}
	\centering
	\includegraphics[width=\textwidth]{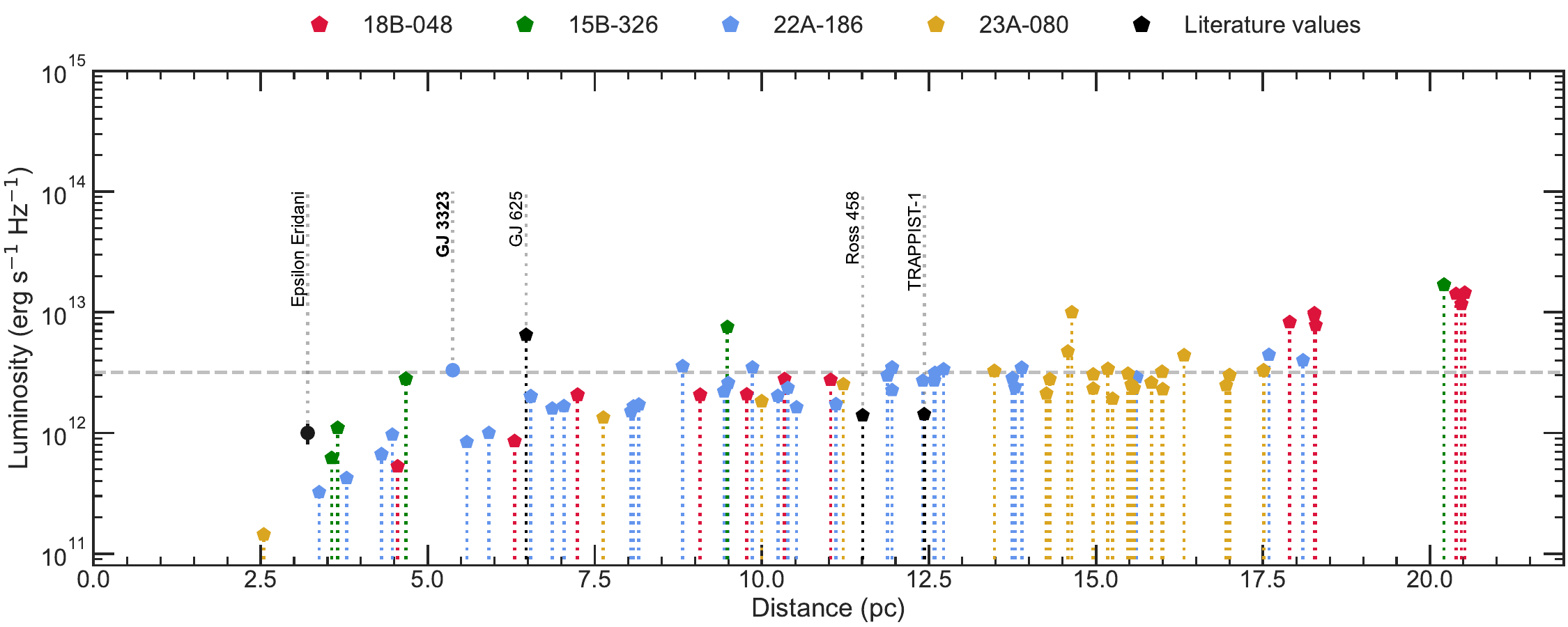}
	\caption{\label{fig:LumLimits} Luminosity upper limits as a function of system distance. A dashed line shows the intended sensitivity of the survey at $L_\nu\lesssim 10^{12.5}$ erg s$^{-1}$ Hz$^{-1}$. Each dataset studied is shown in a different color, and upper limits on luminosity as a function of distance are presented as markers with dotted lines pointing downwards. Results from the literature are also shown for reference; these correspond to the four systems in the unobserved portion of the sample that have published radio observations in the $4-8$ GHz range, taken from \cite{bower_radio_2009, bastian_radio_2018, pineda_deep_2018, cendes_pilot_2021}.}
\end{figure*}

\begin{figure*}
	\centering
	\includegraphics[width=\textwidth]{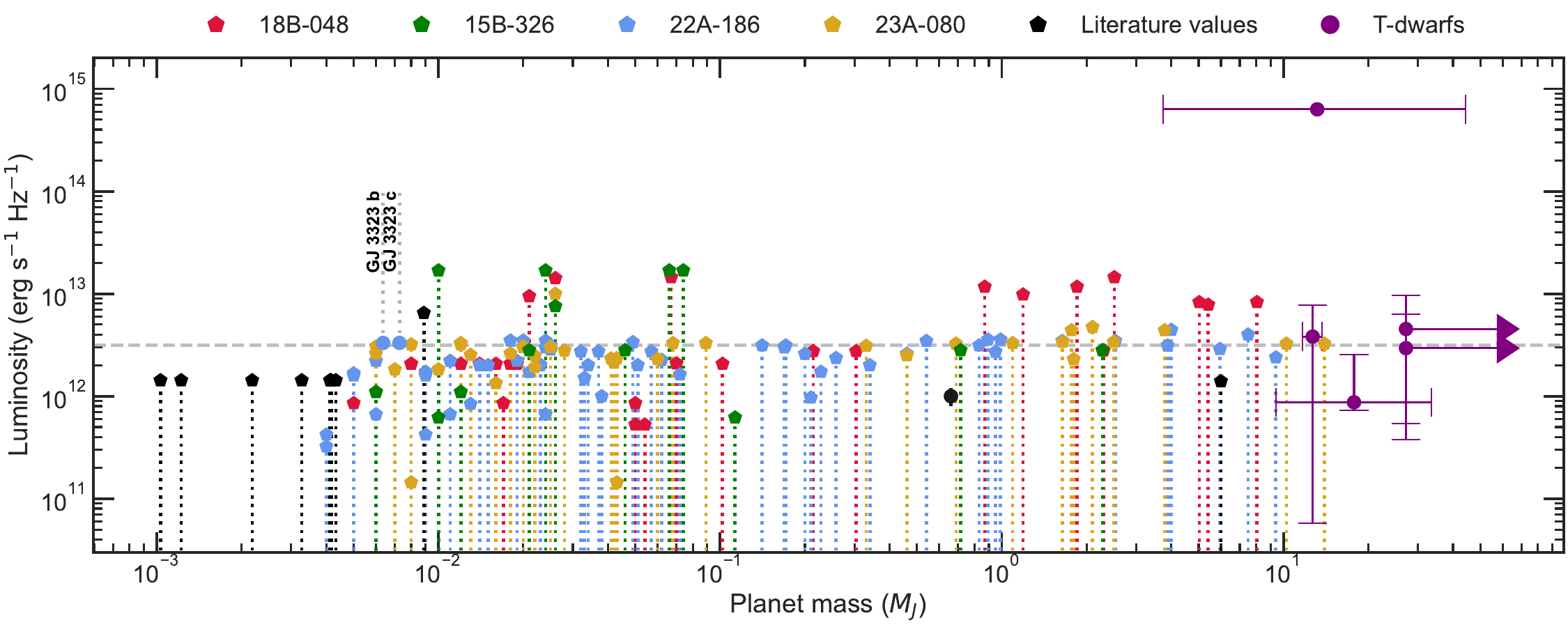}
	\caption{\label{fig:LumLimits_mass} Luminosity limits as a function of planet mass. Here, each planet in the sample is plotted, with the luminosity measurement corresponding to its system. We include the same 4 literature systems as in Figure \ref{fig:LumLimits}. We also include the measured luminosities and estimated masses for the available radio-detected T-dwarfs in the literature: SIMP J013656.5+093347.3, 2MASS J10475385+2124234, 2MASS J12373919+6526148, 2MASS J12545393-0122474 and WISE J062309.94–045624.6. These literature measurements are taken from \cite{kao_auroral_2016, kao_strongest_2018, rose_periodic_2023}}
\end{figure*}

\begin{figure*}
	\centering
	\includegraphics[width=\textwidth]{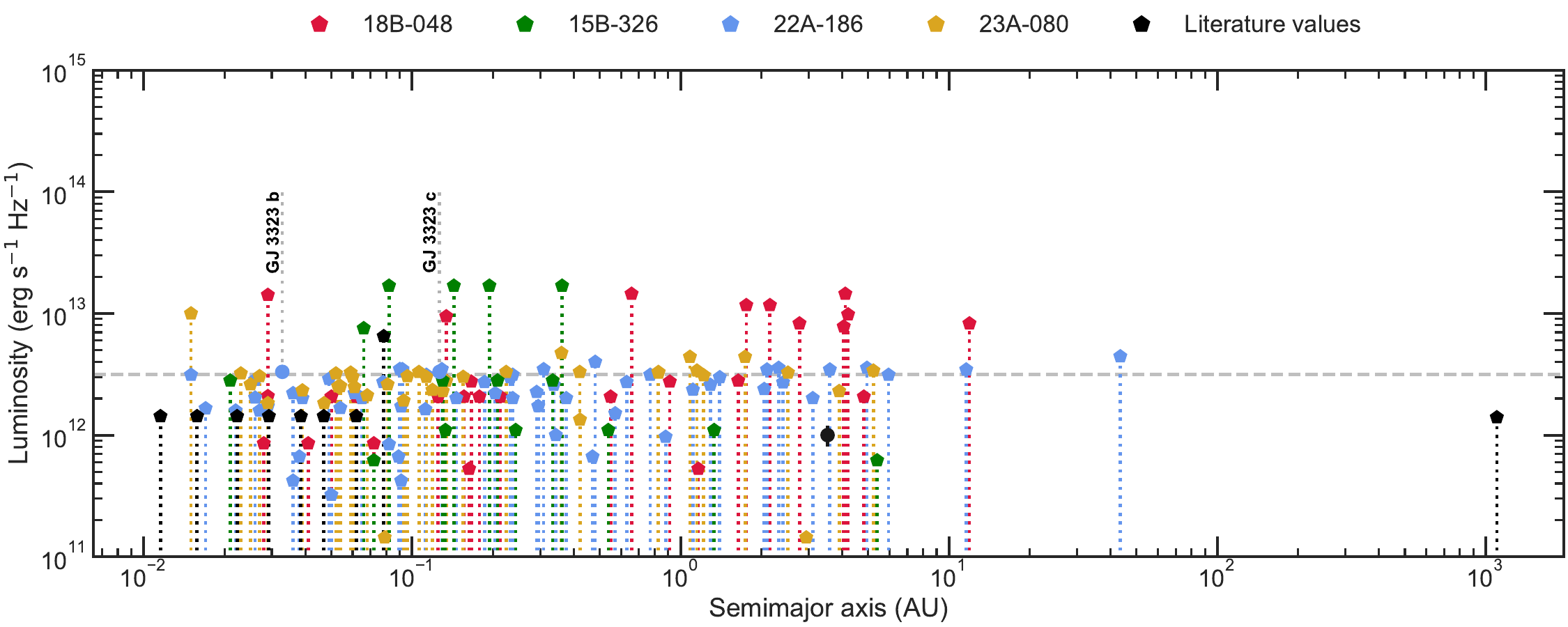}
	\caption{\label{fig:LumLimits_period} Luminosity limits as a function of planet orbital separation. Literature values correspond to the same four systems from Figures \ref{fig:LumLimits} and \ref{fig:LumLimits_mass}.}
\end{figure*}

For programs 18B-048, 22A-186, and 23A-080, the calibrated measurement sets were obtained from the National Radio Astronomy Observatory (NRAO) archive, having been processed by the Common Astronomy Software Application package \citep[CASA;][]{mcmullin_casa_2007} standard 6.4.1 pipeline. In the case of program 15B-326 (PI: Bastian), the calibration files were separately obtained from the NRAO archive and used to calibrate the raw visibilities with CASA 4.3.1.

Images for each target were made using the standard \texttt{CLEAN} algorithm with the CASA \texttt{tclean} procedure, with a pixel size of 1/3 of the synthesized beam size for each observation. We then obtained Gaia DR3 coordinates and proper motions for each target, which have sub-mas and sub-mas/year precision, respectively, for all targets in our sample \citep{gaia_collaboration_gaia_2016, gaia_collaboration_gaia_2023}. We generated proper motion-corrected coordinates for the time of observation for each target.

We used these coordinates to perform point-source photometry on the images at the location of the targets using the \texttt{imtool fitsrc} feature of \texttt{pwkit} \citep{williams_pwkit_2017}. Of the 77 targets, only 1 resulted in a $>5\sigma$ detection of a point source. The resulting flux densities were scaled to spectral luminosities using the distances from the Gaia parallaxes. Results are tabulated in Tables \ref{tab:15}, \ref{tab:18}, \ref{tab:22} and \ref{tab:23}.

\section{Results} 
\label{sec:results}

We obtained one detection and 76 non-detections from the 77 systems, containing 140 exoplanets. The results are shown in Figure~\ref{fig:LumLimits}, where we plot luminosities as a function of distance. At $\lesssim 8$ pc, our luminosity upper limits are $\approx 10^{11}-10^{12}$ erg s$^{-1}$ Hz$^{-1}$, and they reach our nominal target limit of $\approx 10^{12.5}$ erg s$^{-1}$ Hz$^{-1}$ to $\approx 17.5$ pc; the limits beyond 17.5 pc (from archival data) are shallower by about 0.5 dex. Our detection of GJ\,3323 is at a level of $\approx 10^{12.5}$ erg s$^{-1}$ Hz$^{-1}$, and we discuss this in more detail in \S\ref{sec:gj3323}.  These results are consistent with previous searches for radio emission from exoplanet systems at GHz frequencies \citep[e.g.][]{bastian_search_2000, route_rise_2019, cendes_pilot_2021}, which have found no emission from similar targets, although with much smaller sample sizes.

In Figures \ref{fig:LumLimits_mass} and \ref{fig:LumLimits_period} we show the same luminosity limits but now for each exoplanet with respect to their mass and orbital separation, respectively.  Our survey probes a wide planetary mass range of $\approx 10^{-3}-10$ M$_J$. We also compare our results with a few existing measurements of low-mass UCDs for which quiescent radio emission is detected and a mass estimate is available. Unlike planets, for which masses can be measured from their orbital motion, these low-mass stars require comparing observed spectra with atmospheric evolution models to estimate the object's mass.
Finally, Figure \ref{fig:LumLimits_period} shows that we probe orbital separations from  $10^{-2}$ to $10^1$ AU.

In all three Figures we also present a few existing observations from the literature from comparable 4-8 GHz observations.
\cite{pineda_deep_2018} find a limit of $<1.43 \times 10^{12}$ erg s$^{-1}$ Hz$^{-1}$ from $4-8$ GHz observations on TRAPPIST-1.
\cite{bower_radio_2009} found a limit of $<6.5 \times 10^{12}$ erg s$^{-1}$ Hz$^{-1}$ for GJ 625 as part of a survey of stars.
\cite{bastian_radio_2018} detected $\epsilon$ Eridani at $(1.0\pm0.2) \times 10^{12}$ erg s$^{-1}$ Hz$^{-1}$ but conclude that the detection is likely of stellar origin.
We also include the result for the one target in our pilot study \citep{cendes_pilot_2021} that falls within our distance cutoff, Ross 458. That study found a limit of $<1.4 \times 10^{12}$ erg s$^{-1}$ Hz$^{-1}$.
Unlike the limits presented in this work, that limit constrains emission from the planet directly since the planet was resolvable in the observation.
All of these measurements were taken with the VLA.

\begin{figure*}
	\centering
	\includegraphics[width=\textwidth]{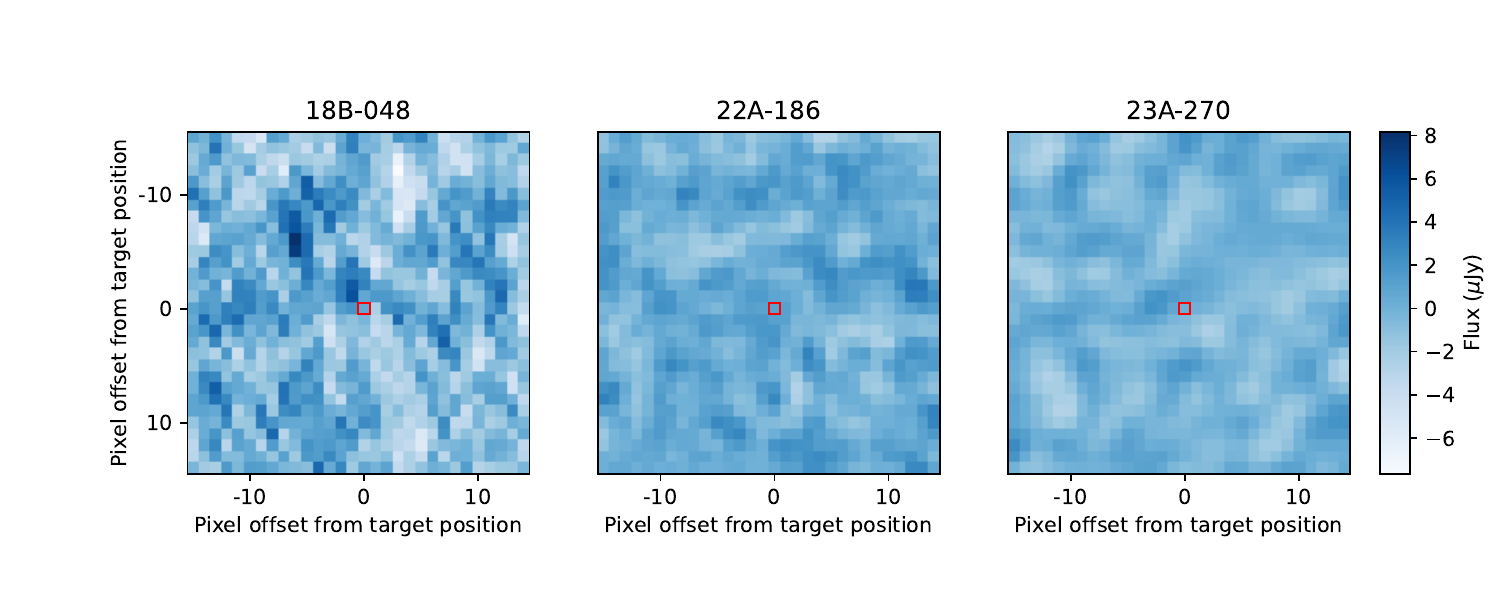}\vspace{-0.2in}
	\caption{\label{fig:stacks} Stacked images for targets from three of the VLA programs reported in this paper. Each stack is made using a weighted average of a $31x31$ pixel region centered on each target star. The center pixel is marked with a red outline. Images were made with a cell size of 1/3 the synthesized beam size, but there may be more than one beam width per stack. The resulting RMS values for the stacks are 2.1 $\mu$Jy (18B-048), 1.1 $\mu$Jy (22A-186), and 1.0 $\mu$Jy (23A-270).
 }
\end{figure*}

Given the individual non-detections, we generated stacked images for each observing program with a sufficient number of targets (i.e., 18B-048, 22A-186, and 23A-270) by aligning the individual images centered on the known position of each source; we stack the images in this manner given the different VLA array configurations (and hence angular resolution) of each program.  In the 22A-186 stack we excluded GJ\,3323 given its individual detection.  The stacked images are shown in Figure~\ref{fig:stacks}, and do not reveal any significant emission at the source locations.  The resulting rms noise levels are 2.1, 1.1 and 1.0 $\mu$Jy for the 18B-048, 22A-186, and 23A-270 stacks, respectively.  
Collectively, this indicates that any steady emission from exoplanets at this frequency range has a typical flux density of $\lesssim 1-2$ $\mu$Jy.

\subsection{Detection of GJ\,3323}
\label{sec:gj3323} 

Our single detection from the survey is of the GJ\,3323 system (5.37 pc), which consists of an M4 star with two terrestrial planets, GJ\,3323~b ($M_p \sin i = 2.02 M_\oplus$, $P = 5.36$ d) and GJ\,3323~c ($M_p \sin i = 2.31 M_\oplus$, $P = 40.54$ d) \citep{astudillo-defru_harps_2017}. GJ\,3323 has been previously detected with the \textit{Chandra X-ray Observatory} with a luminosity of $\log L_X = 27.28$ erg s$^{-1}$ ($0.5 - 8$ keV), and with ROSAT with a luminosity of $\log L_X = 27.45$ erg s$^{-1}$ ($0.1 - 2.4$ keV; \citealt{wright_stellar_2018}). Furthermore, we identify the source in the SRG/eROSITA all-sky survey Data Release 1 \citep{merloni_srgerosita_2024}, with a luminosity of $\log L_X = 27.32$ erg s$^{-1}$ ($0.2 - 2.3$ keV). GJ\,3323 has an estimated Rossby number of 0.87 that places it in the ``unsaturated'' regime of the rotation-activity relation \citep{boudreaux_ca_2022}.

\begin{figure}
	\centering
	\includegraphics[width=\columnwidth]{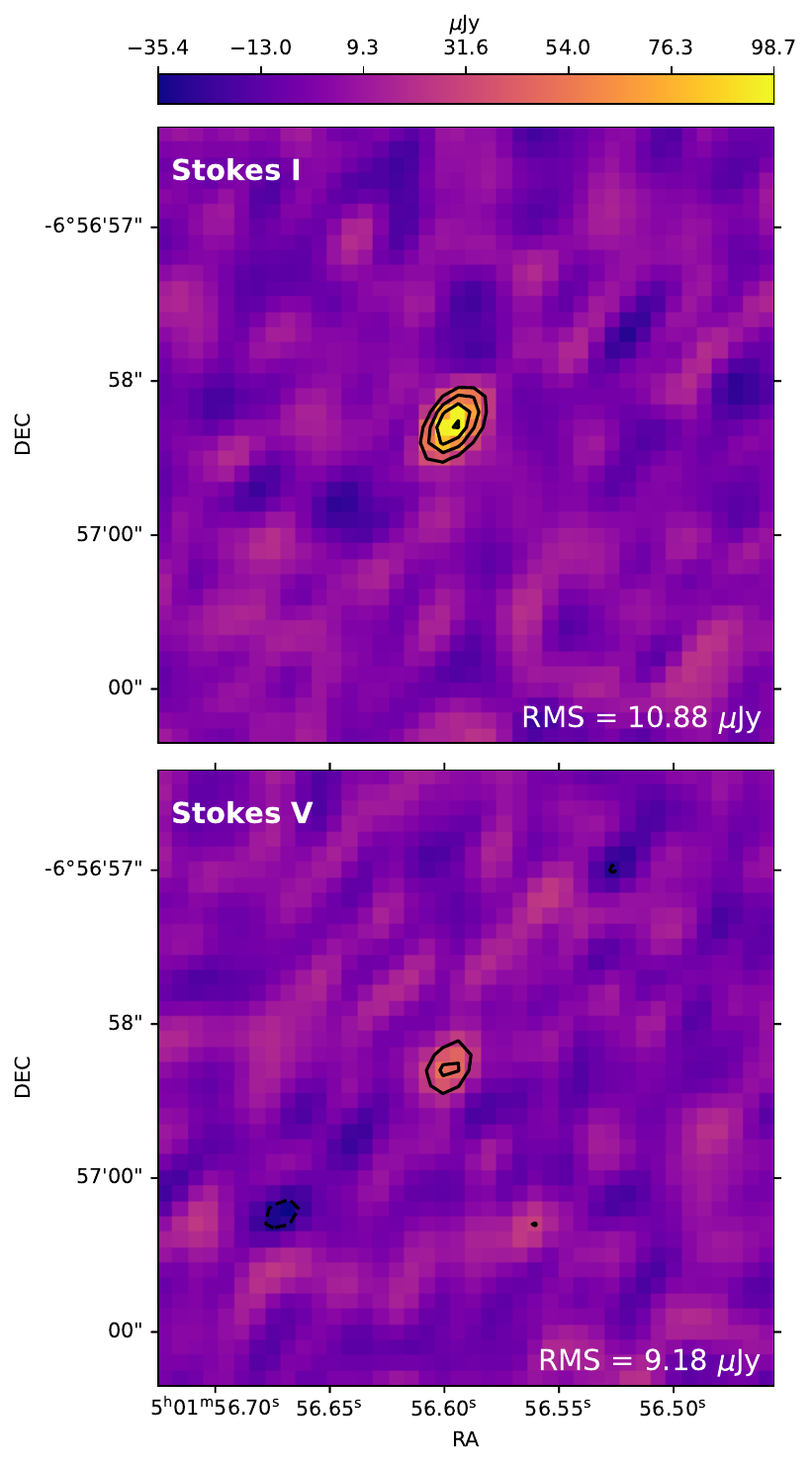}
	\caption{\label{fig:GJ3323} {VLA images of the region centered on the Gaia DR3 proper-motion corrected position of GJ\,3323. The contour levels are $-3,3,5,7,9 \,\sigma$, where $\sigma$ is the RMS of the image as shown in each cutout. GJ\,3323 is detected with a flux density of $86\pm 10$ $\mu$Jy in Stokes I and $35 \pm 9$ $\mu$Jy in Stokes V.}}
\end{figure}

In our VLA observation, we detect GJ\,3323 with a flux density of $86 \pm 10$ $\mu$Jy, corresponding to a luminosity of $\log(L_{\nu}) = 12.47\pm 0.05$ erg s$^{-1}$ Hz$^{-1}$. We also detect it in Stokes V (circular polarization) with a flux density of $35 \pm 9$ $\mu$Jy, corresponding to a circular polarization fraction of $\approx 40$\%. The VLA detection is shown in Figure~\ref{fig:GJ3323}, with the total intensity (Stokes I) in the top panel and the circular polarization (Stokes V) in the bottom panel.

Using the radio and X-ray luminosities, we can compare the results for GJ\,3323 to the G\"udel-Benz Relation (GBR; \citealt{guedel_x-raymicrowave_1993, benz_x-raymicrowave_1994}), an empirical power law relation between the radio and X-ray luminosities of active stars. Stars of spectral types earlier than M7 typically closely follow this relation, spanning almost ten orders of magnitude in radio and X-ray luminosities \citep{williams_radio_2018}. 

\begin{figure}
\centering
\includegraphics[width=\columnwidth]{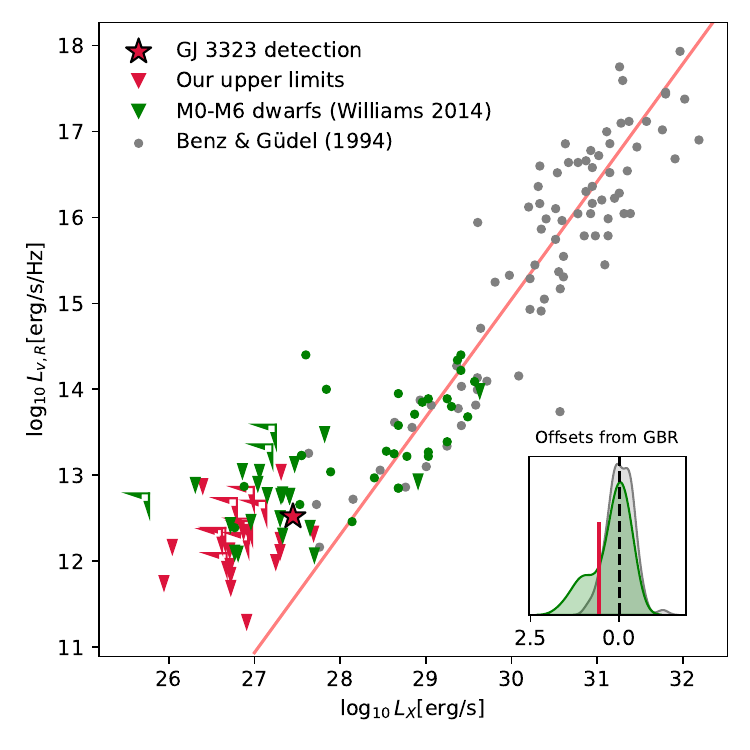}
\caption{\label{fig:gbr} {The G\"udel-Benz Relation (GBR) between X-ray and radio luminosities. The red arrows indicate upper limits on GBR placement obtained from our radio luminosity upper limits, and X-ray luminosity values from \citet{stelzer_uv_2013}. The red star indicates the placement of GJ\,3323 in the GBR from our detection. Gray circles are from the original result of \cite{benz_x-raymicrowave_1994}, and green circles and arrows are detections and upper limits, respectively, of early M-dwarfs (M0-M6) from \cite{williams_trends_2014}. The inset plot shows the distribution of offsets perpendicular to the GBR fit in units of dex for the original Benz-G\"udel sample (grey) and for the \cite{williams_trends_2014} sample (green), with GJ\,3323 as the red line.}}
\end{figure}

We find that GJ\,3323 is located close the GBR, indicating that the radio emission is consistent with having a stellar origin. Our GJ\,3323 detection places it 0.57 dex perpendicular from the GBR best-fit line, while the perpendicular scatter of the original G\"udel-Benz sample is 0.2 dex \citep{williams_trends_2014}. However, stars of spectral type M0--M6 with radio and X-ray detections tend to skew to the left of the GBR fit \citep{williams_trends_2014}, and GJ\,3323 is not exceptional in this (see the inset of Figure \ref{fig:gbr}).

It is important to note that M dwarf X-ray and radio emission can show flaring and variability on a timescale of minutes to hours \citep[e.g.][]{berger_flaring_2002,stelzer_simultaneous_2006,antonova_sporadic_2007}, such that relying on non-simultaneous observations for placing targets in the GBR can be risky. However, the consistent X-ray luminosity from {\it Chandra}, ROSAT, and eROSITA suggests that the X-ray emission is quiescent in nature. For our radio observation, the light curve did not vary over the 11 minute duration, but the short observation time makes further characterization difficult. We also checked VLA Sky Survey \citep[VLASS; ][]{lacy_karl_2020} epochs 1, 2 and 3 for emission from the proper-motion corrected location of GJ 3323 but did not detect a source (to shallow $3\sigma$ limits of $\approx 0.40$ mJy at 3 GHz). 

Despite the overall consistency with a stellar emission origin, the relatively high circular polarization fraction could point to a planetary origin, which we discuss in more detail in \S\ref{sec:discussion}. 

\section{Discussion} 
\label{sec:discussion}

The possibility of radio emission from exoplanet systems has been discussed in the literature in the context of three possible processes: star-planet interaction, direct planetary emission, or stellar emission. We discuss each of these scenarios in turn:

\subsection{Star-planet interaction}

In the solar system, the strength of radio emission from magnetized planets (the radio power output) is directly proportional to the electromagnetic Poynting flux incident on the magnetopause of the planet due to the solar wind, a relation known as the Radiometric Bode's Law (RBL) \citep{desch_predictions_1984}. Historically, the RBL has been used as a scaling law to predict the strength of putative radio emission from exoplanets from their estimated magnetic fields \citep{lazio_radiometric_2004,zarka_plasma_2007}.
However, the RBL is an empirical relation determined only from planets orbiting the same star, our Sun. Given the dependence of this behavior on the solar wind, it is risky to extrapolate this to other stellar systems, especially to systems with stars much different than the Sun.

In the case of M dwarfs, it becomes particularly necessary to take into account that these stars are known to be significantly more active and have a distinct environments from Sun-like stars. Many of these systems also have close-in exoplanets, which have been proposed to be ideal targets for searching for exoplanet-induced radio emission due to increased possibility of observable star-planet interaction stemming from these short orbital separations \citep{cuntz_stellar_2000}. Planets in close orbits around their stars are immersed in flowing magnetized plasma from the stellar environment. The planets themselves thus become obstacles to the plasma flow, and interact with it, causing waves in this flow. In sub-Alfvénic modes, energy gets transported back to the star and can also be observed as radio emission \citep{saur_magnetic_2013}. No solar system planets have this kind of interaction with the Sun, owing to their large orbital distances; sub-Alfvénic interaction is responsible for the observed ``Jupiter-Io'' effect in which periodic radio emission and auroral activity is observed in phase with the orbit of Io \citep{zarka_plasma_2007}, but this is due to magnetospheric currents generated by Jupiter's rotation instead of a wind.

Sub-Alfvénic interaction, however, may be the case in the GJ\,3323 system, as GJ\,3323 b is estimated to be within the Alfvén surface radius of its host star (with GJ\,3323 c just outside the radius) raising the possibility of star-planet interaction as a driver of radio emission \citep{farrish_characterizing_2019}. The radio emission observed from the Jupiter-Io system is coherent and nonthermal, caused by the electron-cyclotron maser instability (ECMI) \citep{zarka_plasma_2007}. In this mechanism, the observed frequency of emission (the cyclotron frequency $v_{\mathrm{c}}$) is proportional to a magnetic field strength and provides a ``point estimate'' of this field strength ($B$) at the point of emission in the object where the cyclotron maser occurs. This means that the field does not need to be this strong everywhere, or even on average, just somewhere in the system. The cyclotron frequency is given by:
\begin{equation} \label{eq:1}
v_{\mathrm{c}}=\frac{e B}{2 \pi m_e c} \approx 2.8\left(\frac{B}{1\, \mathrm{kG}}\right)\,\, \mathrm{GHz}.
\end{equation}
ECMI emission exhibits a sharp drop-off in flux at frequencies higher than the cyclotron frequency, such that the mere detection of ECMI diagnoses the cyclotron frequency and thus the magnetic field strength. For our frequency range of $4-8$ GHz, the above equation yields magnetic field strengths of $1.4-2.8$ kG. It should be noted that the time-dependence of ECMI bursts, as well as significant beaming effects that occur in ECMI emission, introduce additional challenges towards detecting this kind of emission with a short observations like ours \citep{zarka_plasma_2007}.

ECMI emission is typically characterized by a high circular polarization fraction \citep{treumann_electroncyclotron_2006}. The circular polarization fraction of 40\% we detect from GJ\,3323 is unfortunately ambiguous, and especially at GHz frequencies, insufficient to identify the observed emission as caused by ECMI \citep{villadsen_ultra-wideband_2019}. Furthermore, the brief observation presented here cannot truly check for or rule out star-planet interaction since, as a single snapshot, it cannot be correlated with the planets' orbital periods.

\begin{figure}
	\centering
	\includegraphics[width=\columnwidth]{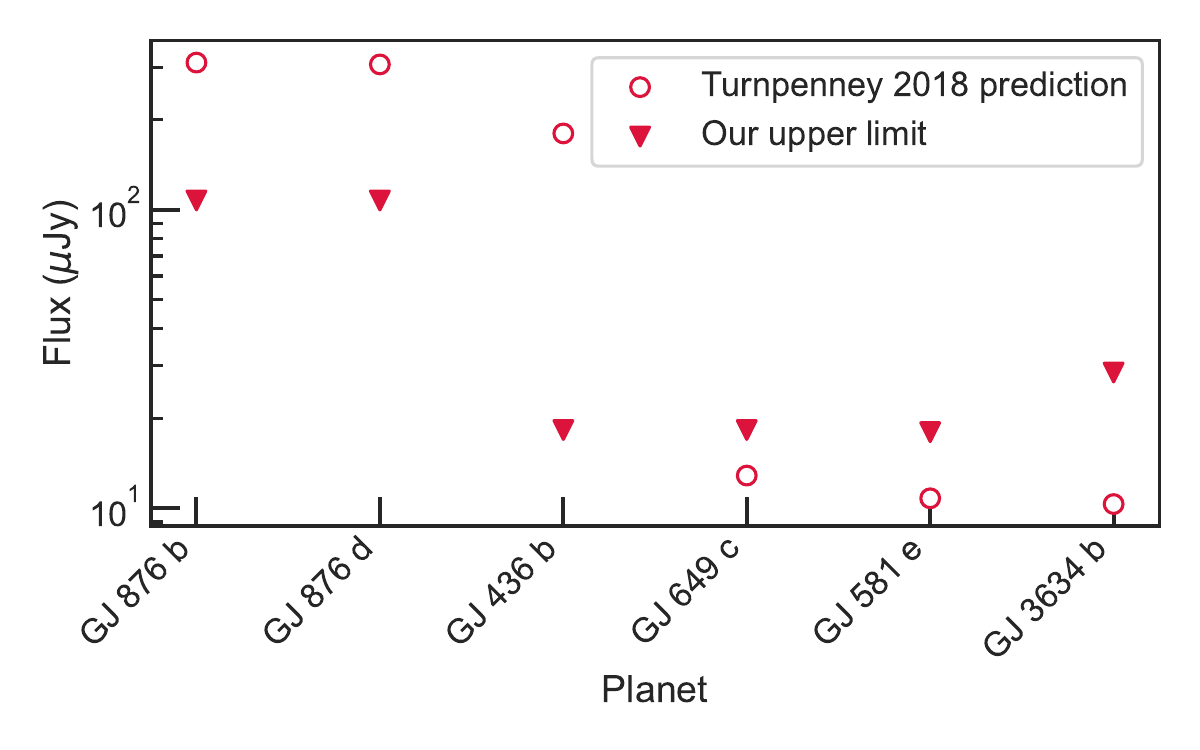}
	\caption{\label{fig:turnpenney_comp} {Flux predictions from \cite{turnpenney_exoplanet-induced_2018} for 6 nearby exoplanets, and our measured upper limits. We establish upper limits more stringent than their predicted fluxes for three planets. Two of these planets belong to the same system, GJ 876.}}
\end{figure}

Beyond the detection of GJ\,3323, we also investigate our upper limits in comparison with existing predictions. \citet{turnpenney_exoplanet-induced_2018} examined the closest M dwarf systems with close-in planets, and modeled this behavior to predict their radio fluxes. In Figure \ref{fig:turnpenney_comp}, we show the predicted fluxes for these systems in comparison to our observed limits. We observed 6 of the 11 exoplanets identified by \citeauthor{turnpenney_exoplanet-induced_2018} as the strongest likely emitters. For three of these planets, our observations establish upper limits that are between 3 and 10 times fainter than the predicted flux. For the remaining three, our limits are about a factor of 2 times higher than the predicted flux.

It is important to note that these predictions involve poorly constrained assumptions about the planetary and stellar magnetic field strengths of the systems in question and the stellar wind mass outflow rates.  Furthermore, the ECMI emission that is considered in this model is taken to have a flat spectral profile up to an unspecified cutoff frequency in the GHz range at which the brightness declines rapidly. The cutoff frequency is proportional to the stellar magnetic field strength in the region of emission. While global magnetic fields for M dwarfs can often reach a few kG \citep{reiners_evidence_2009}, what matters for ECMI emission is the magnetic field strength at the location of emission. Notably, it can plausibly reach the $2-4$ kG threshold probed by our $4-8$ GHz observations even in stars with low global field strength \citep{pineda_deep_2018}.

Recently, \citet{pineda_coherent_2023} published a detection of coherent emission from the YZ Ceti system at $2-4$ GHz using the VLA. Two bursts of emission were detected, and they coincided with the same moment of orbital phase of the only planet in the system, YZ Ceti b, which has a 2 day period orbit. \citet{trigilio_star-planet_2023} independently observed emission in the $0.55- 0.9$ GHz band, using the Giant Metrewave Radio Telescope (GMRT), also in-phase with the planetary orbit. This is tentative evidence that the bursts may be caused by star-planet interaction. In this case, the actual emission may be coming from the star itself, similar to the observed Jupiter-Io effect in the solar system. 

We note that of the observed bursts, one lasted 1 hour and the other lasted only 1 minute. If these signals are in fact the result of star-planet interaction, their occurrence will depend on the planetary orbital period. Given their short duration with respect to a full orbit, then the non-detections presented in this work do not rule out that any of our targets may exhibit these interactions.  The bursts observed from YZ Cet peaked at a luminosity of $L_{\nu} \sim 10^{13}$ erg s$^{-1}$ Hz$^{-1}$, within the sensitivity of our survey.

\subsection{Direct emission from exoplanets}

Beyond emission from star-planet interactions, it is also important to consider direct planetary radio emission. In principle, ECMI emission could be produced and detected directly from an exoplanet. As mentioned previously, if emission were caused by ECMI, a detection at our observed frequencies would correspond to a kilo-Gauss magnetic field; this is beyond the estimated planetary magnetic field strengths of even the largest exoplanets. However, low-mass UCDs were long predicted to have weak magnetic fields \citep{durney_generation_1993, mohanty_activity_2002} before the detection of their bright radio emission.  In UCDs, ECMI emission is observed in flares that can be detected even when the object does not exhibit steady quiescent emission \citep[][]{berger_flaring_2002, route_second_2016}. Furthermore, since the observed cyclotron frequency is proportional to a magnetic field "point estimate", the field does not need to be as strong everywhere, or even on average, just somewhere in the system at a time of observation. 

While convected energy scaling laws suggest that even super-Jupiter exoplanets would exhibit much lower ECMI cyclotron frequencies than the GHz range, the observed UCD emission suggests these scaling laws may not be valid for all planetary-mass objects \citep{christensen_energy_2009, kao_strongest_2018}. On the other hand, models suggest that young or more massive planets could have field strengths as strong as $\sim 0.1$ kG \citep{hori_linkage_2021}; this is still not enough for direct ECMI emission from these planets to be detectable at GHz frequencies.

In addition to the field strength, a population of non-thermal electrons is also required in the planetary environment so that ECMI can take place. These electrons could be provided by the stellar wind, or perhaps by satellites of the planet as occurs in the Jupiter system \citep{noyola_detection_2014,noyola_radio_2016}. Finally, the challenges of the beaming and time-dependence of ECMI bursts mentioned previously also apply, making the prospect of detecting direct emission even more uncertain.

An alternative direct emission mechanism could be gyrosynchrotron emission, which is also present in UCDs in the form of stable, quiescent emission. This type of emission is caused by mildly relativistic electrons moving in a stellar/planetary magnetic field \citep{williams_radio_2018}. Stellar activity could further exacerbate these electrons into producing synchrotron bursts directly in the planetary environment, but this behavior has not yet been observed \citep{gao_observational_2020}. Like with ECMI, both a strong magnetic field and non-thermal electrons are required to be present. While gyrosynchrotron emission can be present at much higher frequencies for a given magnetic field strength compared to ECMI (such that in principle GHz observations could probe weaker magnetic fields than with ECMI observations), as an incoherent mechanism it is also much less efficient, and is expected to be around five orders of magnitude weaker \citep{zarka_magnetospheric_2015}, beyond what can be probed with the sensitivity of the VLA.

\subsection{Stellar Radio Emission} \label{sec:stellar}
 
\begin{figure}
	\centering
	\includegraphics[width=\columnwidth]{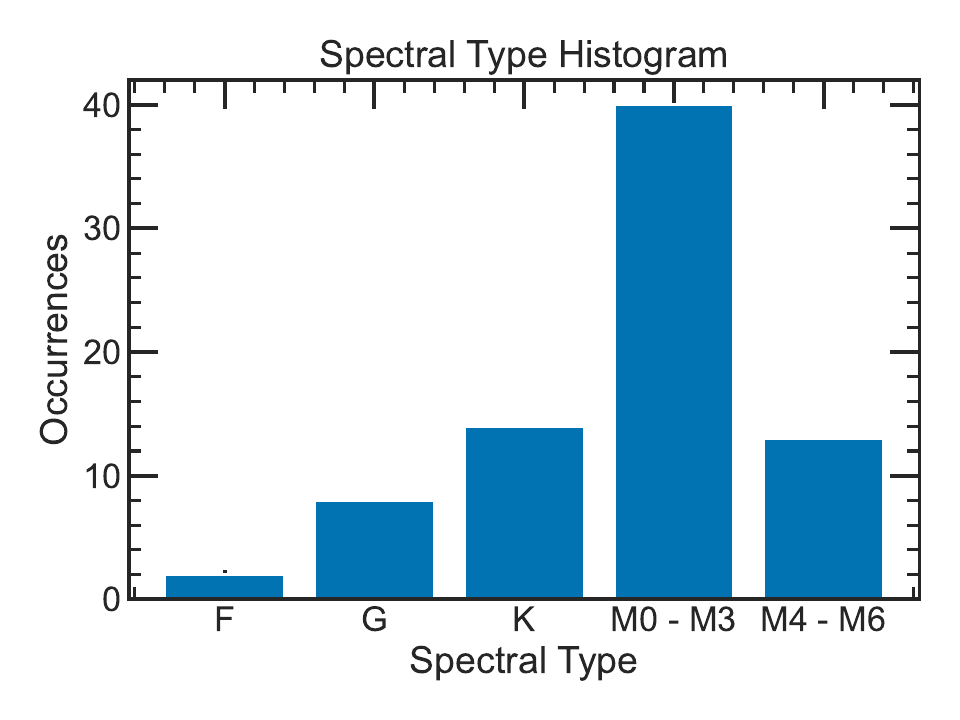}
	\caption{\label{fig:spts} {Spectral type distribution of our observed sample of stellar systems. Spectral type for each observed star is taken from the NASA Exoplanet Archive's Planetary Systems Composite Planet Data Table.}}
\end{figure}

While target selection for our survey was motivated by the known presence of exoplanets, our results are also relevant to the broader study of stellar radio emission. Our target stars span the F, G, K and M spectral types, with the specific breakdown of spectral types shown in Figure \ref{fig:spts}.

Notably, we did not observe any UCDs (spectral types $\gtrsim$ M7); the only two UCDs that meet our survey selection criteria for companion mass, system distance and target declination are TRAPPIST-1 and Teegarden's Star. TRAPPIST-1 has a published luminosity upper limit of $\log_{10}(L_{\nu}) = 12.15$ erg s$^{-1}$ Hz$^{-1}$ from a $4-8$ GHz observation with the VLA \citep{pineda_deep_2018}.  UCDs can be significantly bright in the radio, several orders of magnitude brighter than the GBR would predict \citep{williams_trends_2014}. Meanwhile, earlier type M dwarfs are generally fainter in the radio with respect to their bolometric luminosity, and less likely to be detected at all \citep{berger_radio_2006}.

In our survey, we observed a total of 53 M dwarfs; 40 early M dwarfs (spectral type M0-M3), and 13 mid M dwarfs (M4-M6). Out of these 53 observations, we only detected one star, GJ\,3323. Given the large number of M dwarfs observed, our results are relevant to recent searches for radio activity from these stars \citep[e.g.][]{callingham_population_2021}. It is difficult to gauge the consistency of this survey's results with previous GHz observations of main sequence stars, given differences in sample selection. \citet{bower_radio_2009} surveyed 172 active M dwarfs with the VLA at 5 GHz, and detected 29; their survey sample was built from stars known to be active, for the purpose of identifying bright targets for astrometric study. Our results are more consistent with those of \citet{mclean_radio_2012}, who observed 25 early M dwarfs (M4-M6.5) within 20 pc, detecting only one. However, a systematic study that does not select for activity (or as in our case, the presence of known exoplanets) is necessary to make more definitive conclusions on the radio brightness of these stars.

\section{Conclusions} 
\label{sec:conclusions}

We have presented VLA radio observations at $4-8$ GHz of 77 nearby exoplanet systems, reaching a luminosity sensitivity limit of $\approx 10^{12.5}$ erg s$^{-1}$ Hz$^{-1}$. This sensitivity limit is comparable to our previous pilot study \citep{cendes_pilot_2021} and to detections of radio emission from UCDs (e.g., \citealt{berger_flaring_2002,mclean_radio_2012}), and is more sensitive than previous searches for exoplanet radio emission at GHz frequencies \citep[e.g. ][]{bastian_search_2000}. We detect a single target, GJ\,3323 (M4) with a spectral luminosity of $\log(L_{\nu})\approx 12.5$ erg s$^{-1}$ Hz$^{-1}$. Comparing this result to the known X-ray luminosity of this source, suggests that the emission is likely of stellar origin, but the relatively high fraction of circular polarization may be indicative of star-planet interaction. 

Due to the nature of our survey, observing time was optimized towards reaching a desired sensitivity for a large number of targets. Bursty or intermittent emission may have well been missed in our short observations, although our large number of targets mitigates this limitation in the aggregate, any individual system observed may still be an intermittent emitter. Future long term monitoring of dedicated targets may detect intermittent emission, and may be able to characterize it as of planetary origin through correlation with the planetary orbital period.

Future searches for exoplanet radio emission in the GHz regime may have the capacity to go beyond what has been done in this work thanks to the advent of more sensitive radio telescopes in the next decade, such as the Next-Generation Very Large Array (ngVLA) and the Square Kilometer Array (SKA) \citep{selina_next-generation_2018, braun_anticipated_2019}. It is estimated that SKA1 will achieve an order of magnitude improvement in sensitivity over the VLA for observations of stellar sources, with sensitivity as low as $\sim 2$ $\mu$ Jy for 1 hour integrations. SKA2 and ngVLA will improve another order of magnitude, to $\sim 0.2$ $\mu$Jy \citep{pope_exoplanet_2019}. With these capabilities, it may be possible to either detect or rule out the more optimistic predictions for the brightness of radio emission due to star-planet interactions \citep{turnpenney_exoplanet-induced_2018}.

\begin{acknowledgements}
	We thank Tim Bastian for useful discussions. The Berger Time-Domain Group at Harvard is supported by NSF and NASA grants. The National Radio Astronomy Observatory is a facility of the National Science Foundation operated under cooperative agreement by Associated Universities, Inc. This research has made use of the NASA Exoplanet Archive, which is operated by the California Institute of Technology, under contract with the National Aeronautics and Space Administration under the Exoplanet Exploration Program. We also appreciate the support from the NSF Graduate Research Fellowship (GRFP), grant number DGE1745303, and of the Ford Foundation Predoctoral Fellowship. This work has made use of data from the European Space Agency (ESA) mission {\it Gaia} (\url{https://www.cosmos.esa.int/gaia}), processed by the {\it Gaia} Data Processing and Analysis Consortium (DPAC, \url{https://www.cosmos.esa.int/web/gaia/dpac/consortium}). Funding for the DPAC has been provided by national institutions, in particular the institutions participating in the {\it Gaia} Multilateral Agreement.
\end{acknowledgements}

\bibliographystyle{aasjournal}

\begin{thebibliography}{}
\expandafter\ifx\csname natexlab\endcsname\relax\def\natexlab#1{#1}\fi
\providecommand{\url}[1]{\href{#1}{#1}}
\providecommand{\dodoi}[1]{doi:~\href{http://doi.org/#1}{\nolinkurl{#1}}}
\providecommand{\doeprint}[1]{\href{http://ascl.net/#1}{\nolinkurl{http://ascl.net/#1}}}
\providecommand{\doarXiv}[1]{\href{https://arxiv.org/abs/#1}{\nolinkurl{https://arxiv.org/abs/#1}}}

\bibitem[{Acharya {et~al.}(2023)Acharya, Kashyap, Saar, Singh, \&
  Cuntz}]{acharya_x-ray_2023}
Acharya, A., Kashyap, V.~L., Saar, S.~H., Singh, K.~P., \& Cuntz, M. 2023, The
  Astrophysical Journal, 951, 152, \dodoi{10.3847/1538-4357/acd054}

\bibitem[{Antonova {et~al.}(2007)Antonova, Doyle, Hallinan, Golden, \&
  Koen}]{antonova_sporadic_2007}
Antonova, A., Doyle, J.~G., Hallinan, G., Golden, A., \& Koen, C. 2007,
  Astronomy \& Astrophysics, 472, 257, \dodoi{10.1051/0004-6361:20077231}

\bibitem[{Astudillo-Defru {et~al.}(2017)Astudillo-Defru, Forveille, Bonfils,
  Ségransan, Bouchy, Delfosse, Lovis, Mayor, Murgas, Pepe, Santos, Udry, \&
  Wünsche}]{astudillo-defru_harps_2017}
Astudillo-Defru, N., Forveille, T., Bonfils, X., {et~al.} 2017, Astronomy \&
  Astrophysics, 602, A88, \dodoi{10.1051/0004-6361/201630153}

\bibitem[{Bastian {et~al.}(2000)Bastian, Dulk, \&
  Leblanc}]{bastian_search_2000}
Bastian, T., Dulk, G., \& Leblanc, Y. 2000, The Astrophysical Journal, 545,
  1058

\bibitem[{Bastian {et~al.}(2018)Bastian, Villadsen, Maps, Hallinan, \&
  Beasley}]{bastian_radio_2018}
Bastian, T.~S., Villadsen, J., Maps, A., Hallinan, G., \& Beasley, A.~J. 2018,
  The Astrophysical Journal, 857, 133, \dodoi{10.3847/1538-4357/aab3cb}

\bibitem[{Ben-Jaffel {et~al.}(2021)Ben-Jaffel, Ballester, Muñoz, Lavvas, Sing,
  Sanz-Forcada, Cohen, Kataria, Henry, Buchhave, Mikal-Evans, Wakeford, \&
  López-Morales}]{ben-jaffel_signatures_2021}
Ben-Jaffel, L., Ballester, G.~E., Muñoz, A.~G., {et~al.} 2021, Nature
  Astronomy, 6, 141, \dodoi{10.1038/s41550-021-01505-x}

\bibitem[{Benz \& Guedel(1994)}]{benz_x-raymicrowave_1994}
Benz, A.~O., \& Guedel, M. 1994, Astronomy \& Astrophysics, 285, 621

\bibitem[{Berger(2002)}]{berger_flaring_2002}
Berger, E. 2002, The Astrophysical Journal, 572, 503, \dodoi{10.1086/340301}

\bibitem[{Berger(2006)}]{berger_radio_2006}
---. 2006, The Astrophysical Journal, 648, 629, \dodoi{10.1086/505787}

\bibitem[{Berger {et~al.}(2001)Berger, Ball, Becker, Clarke, Frail, Fukuda,
  Hoffman, Mellon, Momjian, Murphy, Teng, Woodruff, Zauderer, \&
  Zavala}]{berger_discovery_2001}
Berger, E., Ball, S., Becker, K.~M., {et~al.} 2001, Nature, 410, 338,
  \dodoi{10.1038/35066514}

\bibitem[{Berger {et~al.}(2005)Berger, Rutledge, Reid, Bildsten, Gizis,
  Liebert, Martin, Basri, Jayawardhana, Brandeker, Fleming, Johns‐Krull,
  Giampapa, Hawley, \& Schmitt}]{berger_magnetic_2005}
Berger, E., Rutledge, R.~E., Reid, I.~N., {et~al.} 2005, The Astrophysical
  Journal, 627, 960, \dodoi{10.1086/430343}

\bibitem[{Berger {et~al.}(2009)Berger, Rutledge, Phan-Bao, Basri, Giampapa,
  Gizis, Liebert, Martín, \& Fleming}]{berger_periodic_2009}
Berger, E., Rutledge, R.~E., Phan-Bao, N., {et~al.} 2009, The Astrophysical
  Journal, 695, 310, \dodoi{10.1088/0004-637X/695/1/310}

\bibitem[{Blanco-Pozo {et~al.}(2023)Blanco-Pozo, Perger, Damasso,
  Anglada~Escudé, Ribas, Baroch, Caballero, Cifuentes, Jeffers, Lafarga,
  Kaminski, Kaur, Nagel, Perdelwitz, Pérez-Torres, Sozzetti, Viganò, Amado,
  Andreuzzi, Béjar, Brown, Del~Sordo, Dreizler, Galadí-Enríquez, Hatzes,
  Kürster, Lanza, Melis, Molinari, Montes, Murgia, Pallé, Peña-Moñino,
  Perrodin, Pilia, Poretti, Quirrenbach, Reiners, Schweitzer, Zapatero~Osorio,
  \& Zechmeister}]{blanco-pozo_carmenes_2023}
Blanco-Pozo, J., Perger, M., Damasso, M., {et~al.} 2023, Astronomy \&
  Astrophysics, 671, A50, \dodoi{10.1051/0004-6361/202245053}

\bibitem[{Boudreaux {et~al.}(2022)Boudreaux, Newton, Mondrik, Charbonneau, \&
  Irwin}]{boudreaux_ca_2022}
Boudreaux, T.~M., Newton, E.~R., Mondrik, N., Charbonneau, D., \& Irwin, J.
  2022, The Astrophysical Journal, 929, 80, \dodoi{10.3847/1538-4357/ac5cbf}

\bibitem[{Bower {et~al.}(2009)Bower, Bolatto, Ford, \&
  Kalas}]{bower_radio_2009}
Bower, G.~C., Bolatto, A., Ford, E.~B., \& Kalas, P. 2009, The Astrophysical
  Journal, 701, 1922, \dodoi{10.1088/0004-637X/701/2/1922}

\bibitem[{Braun {et~al.}(2019)Braun, Bonaldi, Bourke, Keane, \&
  Wagg}]{braun_anticipated_2019}
Braun, R., Bonaldi, A., Bourke, T., Keane, E., \& Wagg, J. 2019, Anticipated
  {Performance} of the {Square} {Kilometre} {Array} -- {Phase} 1 ({SKA1}),
  \dodoi{10.48550/arXiv.1912.12699}

\bibitem[{Burke \& Franklin(1955)}]{burke_observations_1955}
Burke, B.~F., \& Franklin, K.~L. 1955, Journal of Geophysical Research, 60,
  213, \dodoi{10.1029/JZ060i002p00213}

\bibitem[{Callingham {et~al.}(2021)Callingham, Vedantham, Shimwell, Pope,
  Davis, Best, Hardcastle, Röttgering, Sabater, Tasse, van Weeren, Williams,
  Zarka, de~Gasperin, \& Drabent}]{callingham_population_2021}
Callingham, J.~R., Vedantham, H.~K., Shimwell, T.~W., {et~al.} 2021, Nature
  Astronomy, 5, 1233, \dodoi{10.1038/s41550-021-01483-0}

\bibitem[{Cauley {et~al.}(2019)Cauley, Shkolnik, Llama, \&
  Lanza}]{cauley_magnetic_2019}
Cauley, P.~W., Shkolnik, E.~L., Llama, J., \& Lanza, A.~F. 2019, Nature
  Astronomy, 3, 1128, \dodoi{10.1038/s41550-019-0840-x}

\bibitem[{Cendes {et~al.}(2021)Cendes, Williams, \& Berger}]{cendes_pilot_2021}
Cendes, Y., Williams, P. K.~G., \& Berger, E. 2021, The Astronomical Journal,
  163, 15, \dodoi{10.3847/1538-3881/ac32c8}

\bibitem[{Christensen {et~al.}(2009)Christensen, Holzwarth, \&
  Reiners}]{christensen_energy_2009}
Christensen, U.~R., Holzwarth, V., \& Reiners, A. 2009, Nature, 457, 167,
  \dodoi{10.1038/nature07626}

\bibitem[{Climent {et~al.}(2023)Climent, Guirado, Pérez-Torres, Marcaide, \&
  Peña-Moñino}]{climent_evidence_2023}
Climent, J.~B., Guirado, J.~C., Pérez-Torres, M., Marcaide, J.~M., \&
  Peña-Moñino, L. 2023, Science, 381, 1120, \dodoi{10.1126/science.adg6635}

\bibitem[{Cuntz {et~al.}(2000)Cuntz, Saar, \& Musielak}]{cuntz_stellar_2000}
Cuntz, M., Saar, S.~H., \& Musielak, Z.~E. 2000, The Astrophysical Journal,
  533, L151, \dodoi{10.1086/312609}

\bibitem[{Desch \& Kaiser(1984)}]{desch_predictions_1984}
Desch, M.~D., \& Kaiser, M.~L. 1984, Nature, 310, 755, \dodoi{10.1038/310755a0}

\bibitem[{Driessen {et~al.}(2023)Driessen, Heald, Duchesne, Murphy, Lenc,
  Leung, \& Moss}]{driessen_detection_2023}
Driessen, L.~N., Heald, G., Duchesne, S.~W., {et~al.} 2023, Publications of the
  Astronomical Society of Australia, 40, e036, \dodoi{10.1017/pasa.2023.26}

\bibitem[{Dulk(1985)}]{dulk_radio_1985}
Dulk, G.~A. 1985, Annual review of astronomy and astrophysics, 23, 169

\bibitem[{Durney {et~al.}(1993)Durney, De~Young, \&
  Roxburgh}]{durney_generation_1993}
Durney, B.~R., De~Young, D.~S., \& Roxburgh, I.~W. 1993, Solar Physics, 145,
  207, \dodoi{10.1007/BF00690652}

\bibitem[{Farrish {et~al.}(2019)Farrish, Alexander, Maruo, DeRosa, Toffoletto,
  \& Sciola}]{farrish_characterizing_2019}
Farrish, A.~O., Alexander, D., Maruo, M., {et~al.} 2019, The Astrophysical
  Journal, 885, 51, \dodoi{10.3847/1538-4357/ab4652}

\bibitem[{{Gaia Collaboration} {et~al.}(2016){Gaia Collaboration}, Prusti,
  de~Bruijne, Brown, Vallenari, Babusiaux, Bailer-Jones, Bastian, Biermann,
  Evans, Eyer, Jansen, Jordi, Klioner, Lammers, Lindegren, Luri, Mignard,
  Milligan, Panem, Poinsignon, Pourbaix, Randich, Sarri, Sartoretti, Siddiqui,
  Soubiran, Valette, van Leeuwen, Walton, Aerts, Arenou, Cropper, Drimmel,
  Høg, Katz, Lattanzi, O'Mullane, Grebel, Holland, Huc, Passot, Bramante,
  Cacciari, Castañeda, Chaoul, Cheek, De~Angeli, Fabricius, Guerra,
  Hernández, Jean-Antoine-Piccolo, Masana, Messineo, Mowlavi, Nienartowicz,
  Ordóñez-Blanco, Panuzzo, Portell, Richards, Riello, Seabroke, Tanga,
  Thévenin, Torra, Els, Gracia-Abril, Comoretto, Garcia-Reinaldos, Lock,
  Mercier, Altmann, Andrae, Astraatmadja, Bellas-Velidis, Benson, Berthier,
  Blomme, Busso, Carry, Cellino, Clementini, Cowell, Creevey, Cuypers,
  Davidson, De~Ridder, de~Torres, Delchambre, Dell'Oro, Ducourant, Frémat,
  García-Torres, Gosset, Halbwachs, Hambly, Harrison, Hauser, Hestroffer,
  Hodgkin, Huckle, Hutton, Jasniewicz, Jordan, Kontizas, Korn, Lanzafame,
  Manteiga, Moitinho, Muinonen, Osinde, Pancino, Pauwels, Petit, Recio-Blanco,
  Robin, Sarro, Siopis, Smith, Smith, Sozzetti, Thuillot, van Reeven, Viala,
  Abbas, Abreu~Aramburu, Accart, Aguado, Allan, Allasia, Altavilla, Álvarez,
  Alves, Anderson, Andrei, Anglada~Varela, Antiche, Antoja, Antón, Arcay,
  Atzei, Ayache, Bach, Baker, Balaguer-Núñez, Barache, Barata, Barbier,
  Barblan, Baroni, Barrado~y Navascués, Barros, Barstow, Becciani, Bellazzini,
  Bellei, Bello~García, Belokurov, Bendjoya, Berihuete, Bianchi, Bienaymé,
  Billebaud, Blagorodnova, Blanco-Cuaresma, Boch, Bombrun, Borrachero,
  Bouquillon, Bourda, Bouy, Bragaglia, Breddels, Brouillet, Brüsemeister,
  Bucciarelli, Budnik, Burgess, Burgon, Burlacu, Busonero, Buzzi, Caffau,
  Cambras, Campbell, Cancelliere, Cantat-Gaudin, Carlucci, Carrasco,
  Castellani, Charlot, Charnas, Charvet, Chassat, Chiavassa, Clotet, Cocozza,
  Collins, Collins, Costigan, Crifo, Cross, Crosta, Crowley, Dafonte, Damerdji,
  Dapergolas, David, David, De~Cat, de~Felice, de~Laverny, De~Luise, De~March,
  de~Martino, de~Souza, Debosscher, del Pozo, Delbo, Delgado, Delgado,
  di~Marco, Di~Matteo, Diakite, Distefano, Dolding, Dos~Anjos, Drazinos,
  Durán, Dzigan, Ecale, Edvardsson, Enke, Erdmann, Escolar, Espina, Evans,
  Eynard~Bontemps, Fabre, Fabrizio, Faigler, Falcão, Farràs~Casas, Faye,
  Federici, Fedorets, Fernández-Hernández, Fernique, Fienga, Figueras,
  Filippi, Findeisen, Fonti, Fouesneau, Fraile, Fraser, Fuchs, Furnell, Gai,
  Galleti, Galluccio, Garabato, García-Sedano, Garé, Garofalo, Garralda,
  Gavras, Gerssen, Geyer, Gilmore, Girona, Giuffrida, Gomes, González-Marcos,
  González-Núñez, González-Vidal, Granvik, Guerrier, Guillout, Guiraud,
  Gúrpide, Gutiérrez-Sánchez, Guy, Haigron, Hatzidimitriou, Haywood, Heiter,
  Helmi, Hobbs, Hofmann, Holl, Holland, Hunt, Hypki, Icardi, Irwin, Jevardat~de
  Fombelle, Jofré, Jonker, Jorissen, Julbe, Karampelas, Kochoska, Kohley,
  Kolenberg, Kontizas, Koposov, Kordopatis, Koubsky, Kowalczyk, Krone-Martins,
  Kudryashova, Kull, Bachchan, Lacoste-Seris, Lanza, Lavigne,
  Le~Poncin-Lafitte, Lebreton, Lebzelter, Leccia, Leclerc, Lecoeur-Taibi,
  Lemaitre, Lenhardt, Leroux, Liao, Licata, Lindstrøm, Lister, Livanou, Lobel,
  Löffler, López, Lopez-Lozano, Lorenz, Loureiro, MacDonald,
  Magalhães~Fernandes, Managau, Mann, Mantelet, Marchal, Marchant, Marconi,
  Marie, Marinoni, Marrese, Marschalkó, Marshall, Martín-Fleitas, Martino,
  Mary, Matijevič, Mazeh, McMillan, Messina, Mestre, Michalik, Millar,
  Miranda, Molina, Molinaro, Molinaro, Molnár, Moniez, Montegriffo, Monteiro,
  Mor, Mora, Morbidelli, Morel, Morgenthaler, Morley, Morris, Mulone, Muraveva,
  Musella, Narbonne, Nelemans, Nicastro, Noval, Ordénovic, Ordieres-Meré,
  Osborne, Pagani, Pagano, Pailler, Palacin, Palaversa, Parsons, Paulsen,
  Pecoraro, Pedrosa, Pentikäinen, Pereira, Pichon, Piersimoni, Pineau, Plachy,
  Plum, Poujoulet, Prša, Pulone, Ragaini, Rago, Rambaux, Ramos-Lerate,
  Ranalli, Rauw, Read, Regibo, Renk, Reylé, Ribeiro, Rimoldini, Ripepi, Riva,
  Rixon, Roelens, Romero-Gómez, Rowell, Royer, Rudolph, Ruiz-Dern, Sadowski,
  Sagristà~Sellés, Sahlmann, Salgado, Salguero, Sarasso, Savietto, Schnorhk,
  Schultheis, Sciacca, Segol, Segovia, Segransan, Serpell, Shih, Smareglia,
  Smart, Smith, Solano, Solitro, Sordo, Soria~Nieto, Souchay, Spagna, Spoto,
  Stampa, Steele, Steidelmüller, Stephenson, Stoev, Suess, Süveges, Surdej,
  Szabados, Szegedi-Elek, Tapiador, Taris, Tauran, Taylor, Teixeira, Terrett,
  Tingley, Trager, Turon, Ulla, Utrilla, Valentini, van Elteren, Van~Hemelryck,
  van Leeuwen, Varadi, Vecchiato, Veljanoski, Via, Vicente, Vogt, Voss,
  Votruba, Voutsinas, Walmsley, Weiler, Weingrill, Werner, Wevers, Whitehead,
  Wyrzykowski, Yoldas, Žerjal, Zucker, Zurbach, Zwitter, Alecu, Allen,
  Allende~Prieto, Amorim, Anglada-Escudé, Arsenijevic, Azaz, Balm, Beck,
  Bernstein, Bigot, Bijaoui, Blasco, Bonfigli, Bono, Boudreault, Bressan,
  Brown, Brunet, Bunclark, Buonanno, Butkevich, Carret, Carrion, Chemin,
  Chéreau, Corcione, Darmigny, de~Boer, de~Teodoro, de~Zeeuw, Delle~Luche,
  Domingues, Dubath, Fodor, Frézouls, Fries, Fustes, Fyfe, Gallardo, Gallegos,
  Gardiol, Gebran, Gomboc, Gómez, Grux, Gueguen, Heyrovsky, Hoar, Iannicola,
  Isasi~Parache, Janotto, Joliet, Jonckheere, Keil, Kim, Klagyivik, Klar,
  Knude, Kochukhov, Kolka, Kos, Kutka, Lainey, LeBouquin, Liu, Loreggia,
  Makarov, Marseille, Martayan, Martinez-Rubi, Massart, Meynadier, Mignot,
  Munari, Nguyen, Nordlander, Ocvirk, O'Flaherty, Olias~Sanz, Ortiz, Osorio,
  Oszkiewicz, Ouzounis, Palmer, Park, Pasquato, Peltzer, Peralta, Péturaud,
  Pieniluoma, Pigozzi, Poels, Prat, Prod'homme, Raison, Rebordao, Risquez,
  Rocca-Volmerange, Rosen, Ruiz-Fuertes, Russo, Sembay, Serraller~Vizcaino,
  Short, Siebert, Silva, Sinachopoulos, Slezak, Soffel, Sosnowska, Straižys,
  ter Linden, Terrell, Theil, Tiede, Troisi, Tsalmantza, Tur, Vaccari, Vachier,
  Valles, Van~Hamme, Veltz, Virtanen, Wallut, Wichmann, Wilkinson, Ziaeepour,
  \& Zschocke}]{gaia_collaboration_gaia_2016}
{Gaia Collaboration}, Prusti, T., de~Bruijne, J. H.~J., {et~al.} 2016,
  Astronomy and Astrophysics, 595, A1, \dodoi{10.1051/0004-6361/201629272}

\bibitem[{{Gaia Collaboration} {et~al.}(2023){Gaia Collaboration}, Vallenari,
  Brown, Prusti, de~Bruijne, Arenou, Babusiaux, Biermann, Creevey, Ducourant,
  Evans, Eyer, Guerra, Hutton, Jordi, Klioner, Lammers, Lindegren, Luri,
  Mignard, Panem, Pourbaix, Randich, Sartoretti, Soubiran, Tanga, Walton,
  Bailer-Jones, Bastian, Drimmel, Jansen, Katz, Lattanzi, van Leeuwen, Bakker,
  Cacciari, Castañeda, De~Angeli, Fabricius, Fouesneau, Frémat, Galluccio,
  Guerrier, Heiter, Masana, Messineo, Mowlavi, Nicolas, Nienartowicz, Pailler,
  Panuzzo, Riclet, Roux, Seabroke, Sordo, Thévenin, Gracia-Abril, Portell,
  Teyssier, Altmann, Andrae, Audard, Bellas-Velidis, Benson, Berthier, Blomme,
  Burgess, Busonero, Busso, Cánovas, Carry, Cellino, Cheek, Clementini,
  Damerdji, Davidson, de~Teodoro, Nuñez~Campos, Delchambre, Dell'Oro, Esquej,
  Fernández-Hernández, Fraile, Garabato, García-Lario, Gosset, Haigron,
  Halbwachs, Hambly, Harrison, Hernández, Hestroffer, Hodgkin, Holl, Janßen,
  Jevardat~de Fombelle, Jordan, Krone-Martins, Lanzafame, Löffler, Marchal,
  Marrese, Moitinho, Muinonen, Osborne, Pancino, Pauwels, Recio-Blanco, Reylé,
  Riello, Rimoldini, Roegiers, Rybizki, Sarro, Siopis, Smith, Sozzetti,
  Utrilla, van Leeuwen, Abbas, Ábrahám, Abreu~Aramburu, Aerts, Aguado, Ajaj,
  Aldea-Montero, Altavilla, Álvarez, Alves, Anders, Anderson, Anglada~Varela,
  Antoja, Baines, Baker, Balaguer-Núñez, Balbinot, Balog, Barache, Barbato,
  Barros, Barstow, Bartolomé, Bassilana, Bauchet, Becciani, Bellazzini,
  Berihuete, Bernet, Bertone, Bianchi, Binnenfeld, Blanco-Cuaresma, Blazere,
  Boch, Bombrun, Bossini, Bouquillon, Bragaglia, Bramante, Breedt, Bressan,
  Brouillet, Brugaletta, Bucciarelli, Burlacu, Butkevich, Buzzi, Caffau,
  Cancelliere, Cantat-Gaudin, Carballo, Carlucci, Carnerero, Carrasco,
  Casamiquela, Castellani, Castro-Ginard, Chaoul, Charlot, Chemin, Chiaramida,
  Chiavassa, Chornay, Comoretto, Contursi, Cooper, Cornez, Cowell, Crifo,
  Cropper, Crosta, Crowley, Dafonte, Dapergolas, David, David, de~Laverny,
  De~Luise, De~March, De~Ridder, de~Souza, de~Torres, del Peloso, del Pozo,
  Delbo, Delgado, Delisle, Demouchy, Dharmawardena, Di~Matteo, Diakite, Diener,
  Distefano, Dolding, Edvardsson, Enke, Fabre, Fabrizio, Faigler, Fedorets,
  Fernique, Fienga, Figueras, Fournier, Fouron, Fragkoudi, Gai,
  Garcia-Gutierrez, Garcia-Reinaldos, García-Torres, Garofalo, Gavel, Gavras,
  Gerlach, Geyer, Giacobbe, Gilmore, Girona, Giuffrida, Gomel, Gomez,
  González-Núñez, González-Santamaría, González-Vidal, Granvik, Guillout,
  Guiraud, Gutiérrez-Sánchez, Guy, Hatzidimitriou, Hauser, Haywood, Helmer,
  Helmi, Sarmiento, Hidalgo, Hilger, Hładczuk, Hobbs, Holland, Huckle,
  Jardine, Jasniewicz, Jean-Antoine~Piccolo, Jiménez-Arranz, Jorissen,
  Juaristi~Campillo, Julbe, Karbevska, Kervella, Khanna, Kontizas, Kordopatis,
  Korn, Kóspál, Kostrzewa-Rutkowska, Kruszyńska, Kun, Laizeau, Lambert,
  Lanza, Lasne, Le~Campion, Lebreton, Lebzelter, Leccia, Leclerc,
  Lecoeur-Taibi, Liao, Licata, Lindstrøm, Lister, Livanou, Lobel, Lorca, Loup,
  Madrero~Pardo, Magdaleno~Romeo, Managau, Mann, Manteiga, Marchant, Marconi,
  Marcos, Marcos~Santos, Marín~Pina, Marinoni, Marocco, Marshall, Martin~Polo,
  Martín-Fleitas, Marton, Mary, Masip, Massari, Mastrobuono-Battisti, Mazeh,
  McMillan, Messina, Michalik, Millar, Mints, Molina, Molinaro, Molnár,
  Monari, Monguió, Montegriffo, Montero, Mor, Mora, Morbidelli, Morel, Morris,
  Muraveva, Murphy, Musella, Nagy, Noval, Ocaña, Ogden, Ordenovic, Osinde,
  Pagani, Pagano, Palaversa, Palicio, Pallas-Quintela, Panahi, Payne-Wardenaar,
  Peñalosa~Esteller, Penttilä, Pichon, Piersimoni, Pineau, Plachy, Plum,
  Poggio, Prša, Pulone, Racero, Ragaini, Rainer, Raiteri, Rambaux, Ramos,
  Ramos-Lerate, Re~Fiorentin, Regibo, Richards, Rios~Diaz, Ripepi, Riva, Rix,
  Rixon, Robichon, Robin, Robin, Roelens, Rogues, Rohrbasser, Romero-Gómez,
  Rowell, Royer, Ruz~Mieres, Rybicki, Sadowski, Sáez~Núñez,
  Sagristà~Sellés, Sahlmann, Salguero, Samaras, Sanchez~Gimenez, Sanna,
  Santoveña, Sarasso, Schultheis, Sciacca, Segol, Segovia, Ségransan, Semeux,
  Shahaf, Siddiqui, Siebert, Siltala, Silvelo, Slezak, Slezak, Smart, Snaith,
  Solano, Solitro, Souami, Souchay, Spagna, Spina, Spoto, Steele,
  Steidelmüller, Stephenson, Süveges, Surdej, Szabados, Szegedi-Elek, Taris,
  Taylor, Teixeira, Tolomei, Tonello, Torra, Torra, Torralba~Elipe, Trabucchi,
  Tsounis, Turon, Ulla, Unger, Vaillant, van Dillen, van Reeven, Vanel,
  Vecchiato, Viala, Vicente, Voutsinas, Weiler, Wevers, Wyrzykowski, Yoldas,
  Yvard, Zhao, Zorec, Zucker, \& Zwitter}]{gaia_collaboration_gaia_2023}
{Gaia Collaboration}, Vallenari, A., Brown, A. G.~A., {et~al.} 2023, Astronomy
  and Astrophysics, 674, A1, \dodoi{10.1051/0004-6361/202243940}

\bibitem[{Gao {et~al.}(2020)Gao, Qian, \& Li}]{gao_observational_2020}
Gao, Y., Qian, L., \& Li, D. 2020, The Astrophysical Journal, 895, 22,
  \dodoi{10.3847/1538-4357/ab881b}

\bibitem[{Gloudemans {et~al.}(2023)Gloudemans, Callingham, Duncan, Saxena,
  Harikane, Hill, Zeimann, Röttgering, Hardcastle, Pineda, Shimwell, Smith, \&
  Wagenveld}]{gloudemans_plausible_2023}
Gloudemans, A.~J., Callingham, J.~R., Duncan, K.~J., {et~al.} 2023, Astronomy
  \& Astrophysics, 678, A161, \dodoi{10.1051/0004-6361/202347141}

\bibitem[{Guedel \& Benz(1993)}]{guedel_x-raymicrowave_1993}
Guedel, M., \& Benz, A.~O. 1993, The Astrophysical Journal, 405, L63,
  \dodoi{10.1086/186766}

\bibitem[{Gurdemir {et~al.}(2012)Gurdemir, Redfield, \&
  Cuntz}]{gurdemir_planet-induced_2012}
Gurdemir, L., Redfield, S., \& Cuntz, M. 2012, Publications of the Astronomical
  Society of Australia, 29, 141, \dodoi{10.1071/AS11074}

\bibitem[{Hallinan {et~al.}(2007)Hallinan, Bourke, Lane, Antonova, Zavala,
  Brisken, Boyle, Vrba, Doyle, \& Golden}]{hallinan_periodic_2007}
Hallinan, G., Bourke, S., Lane, C., {et~al.} 2007, The Astrophysical Journal,
  663, L25, \dodoi{10.1086/519790}

\bibitem[{Hallinan {et~al.}(2015)Hallinan, Littlefair, Cotter, Bourke, Harding,
  Pineda, Butler, Golden, Basri, Doyle, Kao, Berdyugina, Kuznetsov, Rupen, \&
  Antonova}]{hallinan_magnetospherically_2015}
Hallinan, G., Littlefair, S.~P., Cotter, G., {et~al.} 2015, Nature, 523, 568,
  \dodoi{10.1038/nature14619}

\bibitem[{Hori(2021)}]{hori_linkage_2021}
Hori, Y. 2021, The Astrophysical Journal, 908, 77,
  \dodoi{10.3847/1538-4357/abd8d1}

\bibitem[{Hughes {et~al.}(2021)Hughes, Boley, Osten, White, \&
  Leacock}]{hughes_unlocking_2021}
Hughes, A.~G., Boley, A.~C., Osten, R.~A., White, J.~A., \& Leacock, M. 2021,
  The Astronomical Journal, 162, 43, \dodoi{10.3847/1538-3881/ac02c3}

\bibitem[{Kao {et~al.}(2016)Kao, Hallinan, Pineda, Escala, Burgasser, Bourke,
  \& Stevenson}]{kao_auroral_2016}
Kao, M.~M., Hallinan, G., Pineda, J.~S., {et~al.} 2016, The Astrophysical
  Journal, 818, 24, \dodoi{10.3847/0004-637X/818/1/24}

\bibitem[{Kao {et~al.}(2018)Kao, Hallinan, Pineda, Stevenson, \&
  Burgasser}]{kao_strongest_2018}
Kao, M.~M., Hallinan, G., Pineda, J.~S., Stevenson, D., \& Burgasser, A. 2018,
  The Astrophysical Journal Supplement Series, 237, 25,
  \dodoi{10.3847/1538-4365/aac2d5}

\bibitem[{Kao {et~al.}(2023)Kao, Mioduszewski, Villadsen, \&
  Shkolnik}]{kao_resolved_2023}
Kao, M.~M., Mioduszewski, A.~J., Villadsen, J., \& Shkolnik, E.~L. 2023,
  Nature, 619, 272, \dodoi{10.1038/s41586-023-06138-w}

\bibitem[{Kao \& Sebastian~Pineda(2022)}]{kao_radio_2022}
Kao, M.~M., \& Sebastian~Pineda, J. 2022, The Astrophysical Journal, 932, 21,
  \dodoi{10.3847/1538-4357/ac660b}

\bibitem[{Lacy {et~al.}(2020)Lacy, Baum, Chandler, Chatterjee, Clarke, Deustua,
  English, Farnes, Gaensler, Gugliucci, Hallinan, Kent, Kimball, Law, Lazio,
  Marvil, Mao, Medlin, Mooley, Murphy, Myers, Osten, Richards, Rosolowsky,
  Rudnick, Schinzel, Sivakoff, Sjouwerman, Taylor, White, Wrobel, Andernach,
  Beasley, Berger, Bhatnager, Birkinshaw, Bower, Brandt, Brown, Burke-Spolaor,
  Butler, Comerford, Demorest, Fu, Giacintucci, Golap, Güth, Hales, Hiriart,
  Hodge, Horesh, Ivezić, Jarvis, Kamble, Kassim, Liu, Loinard, Lyons, Masters,
  Mezcua, Moellenbrock, Mroczkowski, Nyland, O’Dea, O’Sullivan, Peters,
  Radford, Rao, Robnett, Salcido, Shen, Sobotka, Witz, Vaccari, Weeren, Vargas,
  Williams, \& Yoon}]{lacy_karl_2020}
Lacy, M., Baum, S.~A., Chandler, C.~J., {et~al.} 2020, Publications of the
  Astronomical Society of the Pacific, 132, 035001,
  \dodoi{10.1088/1538-3873/ab63eb}

\bibitem[{Lazio {et~al.}(2009)Lazio, Carmichael, Clark, Elkins, Gudmundsen,
  Mott, Szwajkowski, \& Hennig}]{lazio_blind_2009}
Lazio, T. J.~W., Carmichael, S., Clark, J., {et~al.} 2009, The Astronomical
  Journal, 139, 96, \dodoi{10.1088/0004-6256/139/1/96}

\bibitem[{Lazio \& Farrell(2007)}]{lazio_magnetospheric_2007}
Lazio, T. J.~W., \& Farrell, W.~M. 2007, The Astrophysical Journal, 668, 1182,
  \dodoi{10.1086/519730}

\bibitem[{Lazio {et~al.}(2004)Lazio, Farrell, Dietrick, Greenlees, Hogan,
  Jones, \& Hennig}]{lazio_radiometric_2004}
Lazio, T. J.~W., Farrell, W.~M., Dietrick, J., {et~al.} 2004, The Astrophysical
  Journal, 612, 511, \dodoi{10.1086/422449}

\bibitem[{Lecavelier~des Etangs {et~al.}(2011)Lecavelier~des Etangs, Sirothia,
  Gopal-Krishna, \& Zarka}]{lecavelier_des_etangs_gmrt_2011}
Lecavelier~des Etangs, A., Sirothia, S.~K., Gopal-Krishna, \& Zarka, P. 2011,
  Astronomy \& Astrophysics, 533, A50, \dodoi{10.1051/0004-6361/201117330}

\bibitem[{Lecavelier~des Etangs {et~al.}(2013)Lecavelier~des Etangs, Sirothia,
  Gopal-Krishna, \& Zarka}]{lecavelier_des_etangs_hint_2013}
---. 2013, Astronomy \& Astrophysics, 552, A65,
  \dodoi{10.1051/0004-6361/201219789}

\bibitem[{Lynch {et~al.}(2017)Lynch, Murphy, Kaplan, Ireland, \&
  Bell}]{lynch_search_2017}
Lynch, C., Murphy, T., Kaplan, D., Ireland, M., \& Bell, M. 2017, Monthly
  Notices of the Royal Astronomical Society, 467, 3447

\bibitem[{McLean {et~al.}(2012)McLean, Berger, \& Reiners}]{mclean_radio_2012}
McLean, M., Berger, E., \& Reiners, A. 2012, The Astrophysical Journal, 746,
  23, \dodoi{10.1088/0004-637X/746/1/23}

\bibitem[{McMullin {et~al.}(2007)McMullin, Waters, Schiebel, Young, \&
  Golap}]{mcmullin_casa_2007}
McMullin, J.~P., Waters, B., Schiebel, D., Young, W., \& Golap, K. 2007, in
  Astronomical {Society} of the {Pacific} {Conference} {Series}, Vol. 376,
  Astronomical {Data} {Analysis} {Software} and {Systems} {XVI}, ed. R.~A.
  Shaw, F.~Hill, \& D.~J. Bell, 127

\bibitem[{Merloni {et~al.}(2024)Merloni, Lamer, Liu, Ramos-Ceja, Brunner,
  Bulbul, Dennerl, Doroshenko, Freyberg, Friedrich, Gatuzz, Georgakakis,
  Haberl, Igo, Kreykenbohm, Liu, Maitra, Malyali, Mayer, Nandra, Predehl,
  Robrade, Salvato, Sanders, Stewart, Tubín-Arenas, Weber, Wilms, Arcodia,
  Artis, Aschersleben, Avakyan, Aydar, Bahar, Balzer, Becker, Berger, Boller,
  Bornemann, Brüggen, Brusa, Buchner, Burwitz, Camilloni, Clerc, Comparat,
  Coutinho, Czesla, Dannhauer, Dauner, Dauser, Dietl, Dolag, Dwelly, Egg, Ehl,
  Freund, Friedrich, Gaida, Garrel, Ghirardini, Gokus, Grünwald, Grandis,
  Grotova, Gruen, Gueguen, Hämmerich, Hamaus, Hasinger, Haubner, Homan,
  Ider~Chitham, Joseph, Joyce, König, Kaltenbrunner, Khokhriakova, Kink,
  Kirsch, Kluge, Knies, Krippendorf, Krumpe, Kurpas, Li, Liu, Locatelli,
  Lorenz, Müller, Magaudda, Mannes, McCall, Meidinger, Michailidis, Migkas,
  Muñoz-Giraldo, Musiimenta, Nguyen-Dang, Ni, Olechowska, Ota, Pacaud, Pasini,
  Perinati, Pires, Pommranz, Ponti, Poppenhaeger, Pühlhofer, Rau, Reh,
  Reiprich, Roster, Saeedi, Santangelo, Sasaki, Schmitt, Schneider, Schrabback,
  Schuster, Schwope, Seppi, Serim, Shreeram, Sokolova-Lapa, Starck, Stelzer,
  Stierhof, Suleimanov, Tenzer, Traulsen, Trümper, Tsuge, Urrutia, Veronica,
  Waddell, Willer, Wolf, Yeung, Zainab, Zangrandi, Zhang, Zhang, \&
  Zheng}]{merloni_srgerosita_2024}
Merloni, A., Lamer, G., Liu, T., {et~al.} 2024, Astronomy \& Astrophysics, 682,
  A34, \dodoi{10.1051/0004-6361/202347165}

\bibitem[{Mohanty {et~al.}(2002)Mohanty, Basri, Shu, Allard, \&
  Chabrier}]{mohanty_activity_2002}
Mohanty, S., Basri, G., Shu, F., Allard, F., \& Chabrier, G. 2002, The
  Astrophysical Journal, 571, 469, \dodoi{10.1086/339911}

\bibitem[{Noyola {et~al.}(2014)Noyola, Satyal, \&
  Musielak}]{noyola_detection_2014}
Noyola, J.~P., Satyal, S., \& Musielak, Z.~E. 2014, The Astrophysical Journal,
  791, 25

\bibitem[{Noyola {et~al.}(2016)Noyola, Satyal, \& Musielak}]{noyola_radio_2016}
---. 2016, The Astrophysical Journal, 821, 97,
  \dodoi{10.3847/0004-637X/821/2/97}

\bibitem[{O’Gorman {et~al.}(2018)O’Gorman, Coughlan, Vlemmings, Varenius,
  Sirothia, Ray, \& Olofsson}]{ogorman_search_2018}
O’Gorman, E., Coughlan, C.~P., Vlemmings, W., {et~al.} 2018, Astronomy \&
  Astrophysics, 612, A52

\bibitem[{Parker(1955)}]{parker_hydromagnetic_1955}
Parker, E.~N. 1955, The Astrophysical Journal, 122, 293, \dodoi{10.1086/146087}

\bibitem[{Pineda \& Hallinan(2018)}]{pineda_deep_2018}
Pineda, J.~S., \& Hallinan, G. 2018, The Astrophysical Journal, 866, 155,
  \dodoi{10.3847/1538-4357/aae078}

\bibitem[{Pineda {et~al.}(2017)Pineda, Hallinan, \&
  Kao}]{pineda_panchromatic_2017}
Pineda, J.~S., Hallinan, G., \& Kao, M.~M. 2017, The Astrophysical Journal,
  846, 75, \dodoi{10.3847/1538-4357/aa8596}

\bibitem[{Pineda \& Villadsen(2023)}]{pineda_coherent_2023}
Pineda, J.~S., \& Villadsen, J. 2023, Nature Astronomy, 7, 569,
  \dodoi{10.1038/s41550-023-01914-0}

\bibitem[{Pope {et~al.}(2020)Pope, Bedell, Callingham, Vedantham, Snellen,
  Price-Whelan, \& Shimwell}]{pope_no_2020}
Pope, B. J.~S., Bedell, M., Callingham, J.~R., {et~al.} 2020, The Astrophysical
  Journal Letters, 890, L19, \dodoi{10.3847/2041-8213/ab5b99}

\bibitem[{Pope {et~al.}(2019)Pope, Withers, Callingham, \&
  Vogt}]{pope_exoplanet_2019}
Pope, B. J.~S., Withers, P., Callingham, J.~R., \& Vogt, M.~F. 2019, Monthly
  Notices of the Royal Astronomical Society, 484, 648,
  \dodoi{10.1093/mnras/sty3512}

\bibitem[{Reiners {et~al.}(2009)Reiners, Basri, \&
  Browning}]{reiners_evidence_2009}
Reiners, A., Basri, G., \& Browning, M. 2009, The Astrophysical Journal, 692,
  538, \dodoi{10.1088/0004-637X/692/1/538}

\bibitem[{Reiners \& Christensen(2010)}]{reiners_magnetic_2010}
Reiners, A., \& Christensen, U.~R. 2010, Astronomy \& Astrophysics, 522, A13,
  \dodoi{10.1051/0004-6361/201014251}

\bibitem[{Rose {et~al.}(2023)Rose, Pritchard, Murphy, Caleb, Dobie, Driessen,
  Duchesne, Kaplan, Lenc, \& Wang}]{rose_periodic_2023}
Rose, K., Pritchard, J., Murphy, T., {et~al.} 2023, The Astrophysical Journal
  Letters, 951, L43, \dodoi{10.3847/2041-8213/ace188}

\bibitem[{Route(2019)}]{route_rise_2019}
Route, M. 2019, The Astrophysical Journal, 872, 79,
  \dodoi{10.3847/1538-4357/aafc25}

\bibitem[{Route \& Wolszczan(2012)}]{route_arecibo_2012}
Route, M., \& Wolszczan, A. 2012, The Astrophysical Journal, 747, L22,
  \dodoi{10.1088/2041-8205/747/2/L22}

\bibitem[{Route \& Wolszczan(2016)}]{route_second_2016}
---. 2016, The Astrophysical Journal, 830, 85,
  \dodoi{10.3847/0004-637X/830/2/85}

\bibitem[{Route \& Wolszczan(2023)}]{route_rome_2023}
---. 2023, The Astrophysical Journal, 952, 118,
  \dodoi{10.3847/1538-4357/acd9ad}

\bibitem[{Saur {et~al.}(2013)Saur, Grambusch, Duling, Neubauer, \&
  Simon}]{saur_magnetic_2013}
Saur, J., Grambusch, T., Duling, S., Neubauer, F., \& Simon, S. 2013, Astronomy
  \& Astrophysics, 552, A119

\bibitem[{Schreyer {et~al.}(2023)Schreyer, Owen, Spake, Bahroloom, \&
  Di Giampasquale}]{schreyer_using_2023}
Schreyer, E., Owen, J.~E., Spake, J.~J., Bahroloom, Z., \& Di Giampasquale, S.
  2023, Monthly Notices of the Royal Astronomical Society, 527, 5117,
  \dodoi{10.1093/mnras/stad3528}

\bibitem[{Selina {et~al.}(2018)Selina, McKinnon, Beasley, Murphy, Carilli,
  Butler, Clark, Erickson, Grammer, Jackson, Kent, Mason, Morgan, Ojeda,
  Shillue, Sturgis, \& Urbain}]{selina_next-generation_2018}
Selina, R., McKinnon, M., Beasley, A.~J., {et~al.} 2018, in Ground-based and
  {Airborne} {Telescopes} {VII}, ed. R.~Gilmozzi, H.~K. Marshall, \&
  J.~Spyromilio (Austin, United States: SPIE), 55, \dodoi{10.1117/12.2312089}

\bibitem[{Shkolnik {et~al.}(2003)Shkolnik, Walker, \&
  Bohlender}]{shkolnik_evidence_2003}
Shkolnik, E., Walker, G. A.~H., \& Bohlender, D.~A. 2003, The Astrophysical
  Journal, 597, 1092, \dodoi{10.1086/378583}

\bibitem[{Shkolnik {et~al.}(2005)Shkolnik, Walker, Bohlender, Gu, \&
  Kurster}]{shkolnik_hot_2005}
Shkolnik, E., Walker, G. A.~H., Bohlender, D.~A., Gu, P., \& Kurster, M. 2005,
  The Astrophysical Journal, 622, 1075, \dodoi{10.1086/428037}

\bibitem[{Stelzer {et~al.}(2013)Stelzer, Marino, Micela, Lopez-Santiago, \&
  Liefke}]{stelzer_uv_2013}
Stelzer, B., Marino, A., Micela, G., Lopez-Santiago, J., \& Liefke, C. 2013,
  Monthly Notices of the Royal Astronomical Society, 431, 2063,
  \dodoi{10.1093/mnras/stt225}

\bibitem[{Stelzer {et~al.}(2006)Stelzer, Schmitt, Micela, \&
  Liefke}]{stelzer_simultaneous_2006}
Stelzer, B., Schmitt, J. H. M.~M., Micela, G., \& Liefke, C. 2006, Astronomy \&
  Astrophysics, 460, L35, \dodoi{10.1051/0004-6361:20066488}

\bibitem[{Stevenson(2003)}]{stevenson_planetary_2003}
Stevenson, D.~J. 2003, Earth and Planetary Science Letters, 208, 1,
  \dodoi{10.1016/S0012-821X(02)01126-3}

\bibitem[{Treumann(2006)}]{treumann_electroncyclotron_2006}
Treumann, R.~A. 2006, The Astronomy and Astrophysics Review, 13, 229,
  \dodoi{10.1007/s00159-006-0001-y}

\bibitem[{Trigilio {et~al.}(2018)Trigilio, Umana, Cavallaro, Agliozzo, Leto,
  Buemi, Ingallinera, Bufano, \& Riggi}]{trigilio_detection_2018}
Trigilio, C., Umana, G., Cavallaro, F., {et~al.} 2018, Monthly Notices of the
  Royal Astronomical Society, 481, 217, \dodoi{10.1093/mnras/sty2280}

\bibitem[{Trigilio {et~al.}(2023)Trigilio, Biswas, Leto, Umana, Busa,
  Cavallaro, Das, Chandra, Perez-Torres, Wade, Bordiu, Buemi, Bufano,
  Ingallinera, Loru, \& Riggi}]{trigilio_star-planet_2023}
Trigilio, C., Biswas, A., Leto, P., {et~al.} 2023, Star-{Planet} {Interaction}
  at radio wavelengths in {YZ} {Ceti}: {Inferring} planetary magnetic field,
  arXiv.
\newblock \url{http://arxiv.org/abs/2305.00809}

\bibitem[{Turner {et~al.}(2024)Turner, Grießmeier, Zarka, Zhang, \&
  Mauduit}]{turner_follow-up_2024}
Turner, J.~D., Grießmeier, J.-M., Zarka, P., Zhang, X., \& Mauduit, E. 2024,
  Follow-up {LOFAR} observations of the \${\textbackslash}tau\$ {Boötis}
  exoplanetary system, \dodoi{10.48550/arXiv.2403.16392}

\bibitem[{Turner {et~al.}(2021)Turner, Zarka, Grießmeier, Lazio, Cecconi,
  Emilio~Enriquez, Girard, Jayawardhana, Lamy, Nichols, \&
  de~Pater}]{turner_search_2021}
Turner, J.~D., Zarka, P., Grießmeier, J.-M., {et~al.} 2021, Astronomy \&
  Astrophysics, 645, A59, \dodoi{10.1051/0004-6361/201937201}

\bibitem[{Turnpenney {et~al.}(2018)Turnpenney, Nichols, Wynn, \&
  Burleigh}]{turnpenney_exoplanet-induced_2018}
Turnpenney, S., Nichols, J.~D., Wynn, G.~A., \& Burleigh, M.~R. 2018, The
  Astrophysical Journal, 854, 72, \dodoi{10.3847/1538-4357/aaa59c}

\bibitem[{van Leeuwen(2007)}]{van_leeuwen_validation_2007}
van Leeuwen, F. 2007, Astronomy \& Astrophysics, 474, 653,
  \dodoi{10.1051/0004-6361:20078357}

\bibitem[{Vedantham {et~al.}(2020)Vedantham, Callingham, Shimwell, Tasse, Pope,
  Bedell, Snellen, Best, Hardcastle, Haverkorn, Mechev, O’Sullivan,
  Röttgering, \& White}]{vedantham_coherent_2020}
Vedantham, H.~K., Callingham, J.~R., Shimwell, T.~W., {et~al.} 2020, Nature
  Astronomy, 4, 577, \dodoi{10.1038/s41550-020-1011-9}

\bibitem[{Villadsen \& Hallinan(2019)}]{villadsen_ultra-wideband_2019}
Villadsen, J., \& Hallinan, G. 2019, The Astrophysical Journal, 871, 214,
  \dodoi{10.3847/1538-4357/aaf88e}

\bibitem[{Williams(2018)}]{williams_radio_2018}
Williams, P. K.~G. 2018, in Handbook of {Exoplanets}, ed. H.~J. Deeg \& J.~A.
  Belmonte, 171, \dodoi{10.1007/978-3-319-55333-7_171}

\bibitem[{Williams {et~al.}(2013)Williams, Berger, \&
  Zauderer}]{williams_quasi-quiescent_2013}
Williams, P. K.~G., Berger, E., \& Zauderer, B.~A. 2013, The Astrophysical
  Journal, 767, L30, \dodoi{10.1088/2041-8205/767/2/L30}

\bibitem[{Williams {et~al.}(2017)Williams, Clavel, Newton, \&
  Ryzhkov}]{williams_pwkit_2017}
Williams, P. K.~G., Clavel, M., Newton, E., \& Ryzhkov, D. 2017, pwkit:
  {Astronomical} utilities in {Python}

\bibitem[{Williams {et~al.}(2014)Williams, Cook, \&
  Berger}]{williams_trends_2014}
Williams, P. K.~G., Cook, B.~A., \& Berger, E. 2014, The Astrophysical Journal,
  785, 9, \dodoi{10.1088/0004-637X/785/1/9}

\bibitem[{Winglee {et~al.}(1986)Winglee, Dulk, \&
  Bastian}]{winglee_search_1986}
Winglee, R.~M., Dulk, G.~A., \& Bastian, T.~S. 1986, The Astrophysical Journal,
  309, L59

\bibitem[{Wright {et~al.}(2018)Wright, Newton, Williams, Drake, \&
  Yadav}]{wright_stellar_2018}
Wright, N.~J., Newton, E.~R., Williams, P. K.~G., Drake, J.~J., \& Yadav, R.~K.
  2018, Monthly Notices of the Royal Astronomical Society, 479, 2351,
  \dodoi{10.1093/mnras/sty1670}

\bibitem[{Zarka(1998)}]{zarka_auroral_1998}
Zarka, P. 1998, Journal of Geophysical Research: Planets, 103, 20159

\bibitem[{Zarka(2007)}]{zarka_plasma_2007}
---. 2007, Planetary and Space Science, 55, 598

\bibitem[{Zarka {et~al.}(2015)Zarka, Lazio, \&
  Hallinan}]{zarka_magnetospheric_2015}
Zarka, P., Lazio, J., \& Hallinan, G. 2015, Magnetospheric {Radio} {Emissions}
  from {Exoplanets} with the {SKA}, \dodoi{10.22323/1.215.0120}

\bibitem[{Zarka {et~al.}(1997)Zarka, Queinnec, Ryabov, Ryabov, Shevchenko,
  Arkhipov, Rucker, Denis, Gerbault, Dierich, \&
  Rosolen}]{zarka_ground-based_1997}
Zarka, P., Queinnec, J., Ryabov, B.~P., {et~al.} 1997, in Planetary {Radio}
  {Emission} {IV}, 101--127

\bibitem[{Zarka {et~al.}(2012)Zarka, Bougeret, Briand, Cecconi, Falcke, Girard,
  Grießmeier, Hess, Klein-Wolt, Konovalenko, \&
  {others}}]{zarka_planetary_2012}
Zarka, P., Bougeret, J.-L., Briand, C., {et~al.} 2012, Planetary and Space
  Science, 74, 156

\end{thebibliography}

\appendix

\startlongtable
\begin{deluxetable}{lllllllll}
\tablecolumns{9}
\tablecaption{15B-326 Results}
\tablehead{Target System & Distance & Planet &  Planet mass &  Semimajor axis & RMS &  Luminosity \\
 & (pc) & & ($M_J$) & (AU) & ($\mu$Jy) & (erg s$^{-1}$ Hz$^{-1}$)}
\startdata
Gl 15A     & 3.562 & b & 0.010 & 0.072 & 13.6 & $<6.20 \times 10^{11}$ \\ 
           &       & c & 0.113 & 5.400 &      & \\                        
$\tau$ Cet & 3.652 & e & 0.012 & 0.538 & 23.0 & $<1.10 \times 10^{12}$ \\ 
           &       & f & 0.012 & 1.334 &      & \\                        
           &       & g & 0.006 & 0.133 &      & \\                        
           &       & h & 0.006 & 0.243 &      & \\                        
Gl 876     & 4.672 & b & 2.276 & 0.208 & 35.8 & $<2.81 \times 10^{12}$ \\ 
           &       & c & 0.714 & 0.130 &      & \\                        
           &       & d & 0.021 & 0.021 &      & \\                        
           &       & e & 0.046 & 0.334 &      & \\                        
GJ 176     & 9.485 & b & 0.026 & 0.066 & 23.4 & $<7.56 \times 10^{12}$ \\ 
GJ 3293    & 20.21 & b & 0.074 & 0.143 & 11.5 & $<1.69 \times 10^{13}$ \\ 
           &       & c & 0.066 & 0.362 &      & \\                        
           &       & d & 0.024 & 0.194 &      & \\                        
           &       & e & 0.010 & 0.082 &      & \\    
\enddata
\tablecomments{Luminosities determined from 3 times measured RMS (3$\sigma$) and distance.}
\label{tab:15}
\end{deluxetable}

\startlongtable
\begin{deluxetable}{lllllllll}
\tablecolumns{9}
\tablecaption{18B-048 Results}
\tablehead{Target System & Distance & Planet &  Planet mass &  Semimajor axis & RMS &  Luminosity \\
 & (pc) & & ($M_J$) & (AU) & ($\mu$Jy) & (erg s$^{-1}$ Hz$^{-1}$)}
\startdata
Gl 687    & 4.55  & b & 0.054 & 0.163  & 7.1  & $<5.28 \times 10^{11}$ \\ 
          &       & c & 0.050 & 1.165  &      & \\                        
Gl 581    & 6.3   & b & 0.050 & 0.041  & 6.0  & $<8.55 \times 10^{11}$ \\ 
          &       & c & 0.017 & 0.072  &      & \\                        
          &       & e & 0.005 & 0.028  &      & \\                        
Gl 667C   & 7.243 & b & 0.018 & 0.050  & 11.0 & $<2.07 \times 10^{12}$ \\ 
          &       & c & 0.012 & 0.125  &      & \\                        
          &       & e & 0.008 & 0.213  &      & \\                        
          &       & f & 0.008 & 0.156  &      & \\                        
          &       & g & 0.014 & 0.549  &      & \\                        
Gl 433    & 9.077 & b & 0.019 & 0.062  & 7.0  & $<2.07 \times 10^{12}$ \\ 
          &       & c & 0.102 & 4.819  &      & \\                        
          &       & d & 0.016 & 0.178  &      & \\                        
Gl 436    & 9.775 & b & 0.070 & 0.029  & 6.1  & $<2.09 \times 10^{12}$ \\ 
Pollux\footnote{Target coordinates, proper motion and distance taken from the \textit{Hipparcos} catalogue \citep{van_leeuwen_validation_2007} due to unavailability in \textit{Gaia}.}    & 10.34 & b & 2.300 & 1.640  & 7.3  & $<2.80 \times 10^{12}$ \\ 
HIP 57050 & 11.03 & b & 0.304 & 0.166  & 6.3  & $<2.75 \times 10^{12}$ \\ 
          &       & c & 0.214 & 0.912  &      & \\                        
14 Her    & 17.9  & b & 8.053 & 2.774  & 7.2  & $<8.28 \times 10^{12}$ \\ 
          &       & c & 5.025 & 11.924 &      & \\                        
HD 154088 & 18.27 & b & 0.021 & 0.134  & 7.9  & $<9.47 \times 10^{12}$ \\ 
HD 154345 & 18.27 & b & 1.190 & 4.210  & 8.2  & $<9.82 \times 10^{12}$ \\ 
HD 87883  & 18.29 & b & 5.409 & 4.055  & 6.5  & $<7.81 \times 10^{12}$ \\ 
Gl 3634   & 20.39 & b & 0.026 & 0.029  & 9.5  & $<1.42 \times 10^{13}$ \\ 
7 CMa     & 20.47 & b & 1.850 & 1.758  & 7.8  & $<1.17 \times 10^{13}$ \\ 
          &       & c & 0.870 & 2.153  &      & \\                        
Gl 328    & 20.52 & b & 2.510 & 4.110  & 9.6  & $<1.45 \times 10^{13}$ \\ 
          &       & c & 0.067 & 0.657  &      & \\  
\enddata
\label{tab:18}
\end{deluxetable}

\startlongtable
\begin{deluxetable}{lllllllll}
\tablecolumns{9}
\tablecaption{22A-186 Results}
\tablehead{Target System & Distance & Planet &  Planet mass &  Semimajor axis & RMS &  Luminosity \\
 & (pc) & & ($M_J$) & (AU) & ($\mu$Jy) & (erg s$^{-1}$ Hz$^{-1}$)}
\startdata
Ross 128     & 3.375 & b & 0.004 & 0.050  & 7.9  & $<3.23 \times 10^{11}$ \\
GJ 273       & 3.786 & b & 0.009 & 0.091  & 8.2  & $<4.22 \times 10^{11}$ \\
             &       & c & 0.004 & 0.036  &      & \\
Wolf 1061    & 4.308 & b & 0.006 & 0.038  & 10.0 & $<6.66 \times 10^{11}$ \\
             &       & c & 0.011 & 0.089  &      & \\
             &       & d & 0.024 & 0.470  &      & \\
GJ 9066      & 4.47  & c & 0.210 & 0.880  & 13.5 & $<9.68 \times 10^{11}$ \\
GJ 3323      & 5.375 & b & 0.006 & 0.033  & 7.2  & $3.31 \times 10^{11}$ \\
             &       & c & 0.007 & 0.126  &      & \\
GJ 251       & 5.585 & b & 0.013 & 0.082  & 7.5  & $<8.40 \times 10^{11}$ \\
HD 180617    & 5.915 & b & 0.038 & 0.343  & 8.0  & $<1.00 \times 10^{12}$ \\
HD 219134    & 6.542 & b & 0.015 & 0.039  & 13.1 & $<2.01 \times 10^{12}$ \\
             &       & c & 0.014 & 0.065  &      & \\
             &       & d & 0.051 & 0.237  &      & \\
             &       & f & 0.023 & 0.146  &      & \\
             &       & g & 0.034 & 0.375  &      & \\
             &       & h & 0.340 & 3.110  &      & \\
LTT 1445A\footnote{Target coordinates, proper motion and distance taken from the \textit{Hipparcos} catalogue \citep{van_leeuwen_validation_2007} due to unavailability in \textit{Gaia}.}        & 6.864 & b & 0.009 & 0.022  & 9.4  & $<1.59 \times 10^{12}$ \\
             &       & c & 0.005 & 0.027  &      & \\
GJ 393       & 7.038 & b & 0.005 & 0.054  & 9.4  & $<1.67 \times 10^{12}$ \\
GJ 1151      & 8.043 & c & 0.033 & 0.571  & 6.5  & $<1.51 \times 10^{12}$ \\
GJ 486       & 8.079 & b & 0.009 & 0.017  & 7.1  & $<1.66 \times 10^{12}$ \\
Gl 686       & 8.16  & b & 0.021 & 0.091  & 7.2  & $<1.72 \times 10^{12}$ \\
GJ 849       & 8.815 & b & 0.893 & 2.320  & 12.8 & $<3.57 \times 10^{12}$ \\
             &       & c & 0.990 & 4.950  &      & \\
GJ 357       & 9.436 & b & 0.006 & 0.036  & 6.9  & $<2.21 \times 10^{12}$ \\
             &       & c & 0.011 & 0.061  &      & \\
             &       & d & 0.019 & 0.204  &      & \\
GJ 3512      & 9.497 & b & 0.460 & 0.337  & 8.0  & $<2.59 \times 10^{12}$ \\
             &       & c & 0.200 & 1.292  &      & \\
Gl 49        & 9.86  & b & 0.018 & 0.090  & 10.0 & $<3.49 \times 10^{12}$ \\
GJ 1265      & 10.24 & b & 0.023 & 0.026  & 5.4  & $<2.03 \times 10^{12}$ \\
GJ 649       & 10.39 & b & 0.258 & 1.112  & 6.1  & $<2.36 \times 10^{12}$ \\
HIP 48714    & 10.52 & b & 0.072 & 0.112  & 4.1  & $<1.63 \times 10^{12}$ \\
GJ 740       & 11.11 & b & 0.009 & 0.029  & 3.9  & $<1.73 \times 10^{12}$ \\
HD 3651      & 11.11 & b & 0.228 & 0.295  & 3.9  & $<1.73 \times 10^{12}$ \\
GJ 414A      & 11.88 & b & 0.024 & 0.232  & 5.9  & $<2.99 \times 10^{12}$ \\
             &       & c & 0.169 & 1.400  &      & \\
GJ 180       & 11.95 & b & 0.020 & 0.092  & 6.8  & $<3.48 \times 10^{12}$ \\
GJ 96        & 11.95 & b & 0.062 & 0.291  & 4.4  & $<2.26 \times 10^{12}$ \\
             &       & c & 0.020 & 0.129  &      & \\
             &       & d & 0.024 & 0.309  &      & \\
GJ 179       & 12.41 & b & 0.950 & 2.410  & 4.9  & $<2.71 \times 10^{12}$ \\
HD 69830     & 12.58 & b & 0.032 & 0.078  & 4.8  & $<2.73 \times 10^{12}$ \\
             &       & c & 0.037 & 0.186  &      & \\
             &       & d & 0.057 & 0.630  &      & \\
55 Cancri    & 12.59 & b & 0.831 & 0.113  & 5.5  & $<3.13 \times 10^{12}$ \\
             &       & c & 0.171 & 0.237  &      & \\
             &       & d & 3.878 & 5.957  &      & \\
             &       & e & 0.025 & 0.015  &      & \\
             &       & f & 0.141 & 0.771  &      & \\
HD 190007    & 12.72 & b & 0.049 & 0.092  & 5.8  & $<3.37 \times 10^{12}$ \\
GJ 3779      & 13.75 & b & 0.025 & 0.026  & 4.2  & $<2.85 \times 10^{12}$ \\
$\gamma$ Cep & 13.79 & b & 9.400 & 2.050  & 3.5  & $<2.39 \times 10^{12}$ \\
47 UMa       & 13.89 & b & 2.530 & 2.100  & 5.0  & $<3.46 \times 10^{12}$ \\
             &       & c & 0.540 & 3.600  &      & \\
             &       & d & 1.640 & 11.600 &      & \\
$\tau$ Boo   & 15.61 & b & 5.950 & 0.049  & 3.3  & $<2.89 \times 10^{12}$ \\
GJ 504       & 17.59 & b & 4.000 & 43.500 & 4.0  & $<4.44 \times 10^{12}$ \\
70 Vir       & 18.1  & b & 7.490 & 0.481  & 3.4  & $<4.00 \times 10^{12}$ \\
\enddata
\label{tab:22}
\end{deluxetable}

\startlongtable
\begin{deluxetable}{lllllllll}
\tablecolumns{9}
\tablecaption{23A-080 Results}
\tablehead{Target System & Distance & Planet &  Planet mass &  Semimajor axis & RMS &  Luminosity \\
 & (pc) & & ($M_J$) & (AU) & ($\mu$Jy) & (erg s$^{-1}$ Hz$^{-1}$)}
\startdata
GJ 411         & 2.546 & b & 0.008  & 0.079 & 6.2  & $<1.44 \times 10^{11}$ \\
               &       & c & 0.043  & 2.940 &      & \\
GJ 514         & 7.628 & b & 0.016  & 0.422 & 6.4  & $<1.34 \times 10^{12}$ \\
HD 260655      & 9.998 & b & 0.007  & 0.029 & 5.1  & $<1.83 \times 10^{12}$ \\
               &       & c & 0.010  & 0.047 &      & \\
Ross 508       & 11.22 & b & 0.013  & 0.054 & 5.6  & $<2.53 \times 10^{12}$ \\
$\upsilon$ And & 13.48 & c & 13.980 & 0.828 & 5.0  & $<3.26 \times 10^{12}$ \\
               &       & d & 10.250 & 2.513 &      & \\
               &       & b & 0.688  & 0.059 &      & \\
GJ 480         & 14.26 & b & 0.042  & 0.068 & 2.9  & $<2.12 \times 10^{12}$ \\
GJ 685         & 14.31 & b & 0.028  & 0.134 & 3.8  & $<2.79 \times 10^{12}$ \\
HIP 79431      & 14.58 & b & 2.100  & 0.360 & 6.2  & $<4.73 \times 10^{12}$ \\
GJ 1214        & 14.64 & b & 0.026  & 0.015 & 13.0 & $<1.00 \times 10^{13}$ \\
LHS 1140       & 14.96 & b & 0.020  & 0.096 & 3.8  & $<3.05 \times 10^{12}$ \\
Gl 378         & 14.96 & b & 0.041  & 0.039 & 2.9  & $<2.33 \times 10^{12}$ \\
               &       & c & 0.006  & 0.027 &      & \\
GJ 317         & 15.18 & c & 1.644  & 5.230 & 4.1  & $<3.39 \times 10^{12}$ \\
               &       & b & 2.500  & 1.151 &      & \\
HD 238090      & 15.25 & b & 0.022  & 0.093 & 2.3  & $<1.92 \times 10^{12}$ \\
TYC 2187-512-1 & 15.48 & b & 0.330  & 1.220 & 3.6  & $<3.10 \times 10^{12}$ \\
51 Peg         & 15.53 & b & 0.460  & 0.053 & 2.9  & $<2.51 \times 10^{12}$ \\
GJ 720A        & 15.57 & b & 0.043  & 0.119 & 2.7  & $<2.35 \times 10^{12}$ \\
GJ 3929        & 15.83 & b & 0.006  & 0.025 & 2.9  & $<2.61 \times 10^{12}$ \\
               &       & c & 0.018  & 0.081 &      & \\
G 264-12       & 15.99 & b & 0.008  & 0.023 & 3.5  & $<3.21 \times 10^{12}$ \\
               &       & c & 0.012  & 0.052 &      & \\
HD 190360      & 16.0  & b & 1.800  & 3.900 & 2.5  & $<2.30 \times 10^{12}$ \\
               &       & c & 0.060  & 0.130 &      & \\
HD 128311      & 16.32 & b & 1.769  & 1.084 & 4.6  & $<4.40 \times 10^{12}$ \\
               &       & c & 3.789  & 1.740 &      & \\
GJ 3942        & 16.95 & b & 0.022  & 0.061 & 2.4  & $<2.48 \times 10^{12}$ \\
HD 7924        & 17.0  & d & 0.020  & 0.155 & 2.9  & $<3.01 \times 10^{12}$ \\
               &       & c & 0.025  & 0.113 &      & \\
               &       & b & 0.020  & 0.060 &      & \\
$\rho$ CrB     & 17.51 & b & 1.093  & 0.224 & 3.0  & $<3.30 \times 10^{12}$ \\
               &       & c & 0.089  & 0.421 &      & \\
               &       & d & 0.068  & 0.827 &      & \\
               &       & e & 0.012  & 0.106 &      & \\                       
\enddata
\label{tab:23}
\end{deluxetable}

\end{document}